\documentclass[acmsmall,screen]{acmart}

\AtBeginDocument{%
  }

\renewcommand\footnotetextcopyrightpermission[1]{}
\usepackage{tabularray}
\usepackage[noend]{algpseudocode}
\usepackage{multirow}
\usepackage{xurl}
\usepackage{wrapfig} 
\usepackage{booktabs} 
\usepackage{amsmath}
\usepackage{float}
\usepackage{graphicx}
\usepackage[belowskip=-12pt]{caption}
\usepackage{subcaption}
\usepackage{rotating}
\usepackage{tablefootnote}
\usepackage{adjustbox}
\usepackage{array}
 \usepackage{float}
\usepackage{tabularx}
\usepackage[dvipsnames]{xcolor}
\usepackage{stfloats}
\usepackage{wrapfig}
\usepackage{listings}
\usepackage{color, colortbl}
\usepackage{balance} 
\usepackage{lscape}
\usepackage{xspace}
\usepackage{framed}
\usepackage{syntax}
\usepackage{parcolumns}
\usepackage{pifont}
\usepackage{soul}
\usepackage[tikz]{bclogo}
\usepackage{enumitem}
\usepackage{soul}
\usepackage{makecell}
\usepackage[many]{tcolorbox}
\usepackage{cleveref}
\usepackage{tikzsymbols}
\usepackage{textcomp}
\usepackage{hyperref}
\usepackage{graphicx}
\usepackage{textcomp}
\usepackage{xcolor}
\usepackage{microtype}
\usepackage{wrapfig}
\usepackage{tikz}

\crefformat{section}{\S#2#1#3} 
\crefformat{subsection}{\S#2#1#3}
\crefformat{subsubsection}{\S#2#1#3}

\definecolor{maroon}{cmyk}{0,0.87,0.68,0.00}
\definecolor{mygreen}{cmyk}{0.41,0.1,0.5,0}

\newcolumntype{R}[2]{%
    >{\adjustbox{angle=#1,lap=\width-(#2)}\bgroup}%
    l%
    <{\egroup}%
}

\usepackage[linesnumbered,ruled,vlined]{algorithm2e}
\SetKwInput{KwInputs}{Inputs}                
\SetKwInput{KwOutputs}{Outputs}              
\SetKwInput{KwOutput}{Output}              

\definecolor{patchgreen}{HTML}{ccffcc}
\definecolor{textgreen}{HTML}{66cc66}
\definecolor{patchblue}{HTML}{BDEDFF}
\definecolor{textblue}{HTML}{5ABBDC}
\definecolor{patchred}{HTML}{ffcccc}
\definecolor{textred}{HTML}{ff7777}
\definecolor{myyellow}{HTML}{FFFFCC}
\definecolor{mygray}{HTML}{F0F0F0}

\newcommand{\addgreen}{\makebox[0pt][l]{\color{patchgreen}\rule[-1ex]{\linewidth}{3ex}}}
\newcommand{\addred}{\makebox[0pt][l]{\color{patchred}\rule[-1ex]{\linewidth}{3ex}}}
\newcommand{\addblue}{\makebox[0pt][l]{\color{patchblue}\rule[-1ex]{\linewidth}{3ex}}}

\newcommand{\yang}[1]{\textcolor{blue}{[#1]}}

\newcommand{\reyhan}[1]{\textcolor{purple}{[Reyhan: #1]}}
\newcommand\printpercent[2]{\the\numexpr#1*100/#2\%}

\preto\tabular{\setcounter{rownumbers}{0}}
\newcounter{rownumbers}

\newcommand{\approach}{\textsc{GeneBench}\xspace}
\newcommand{\avatar}{Avatar\xspace}
\newcommand{\cruxeval}{CRUXEval\xspace}
\newcommand{\humaneval}{HumanEval\xspace}
\newcommand{\classeval}{ClassEval\xspace}

\newcommand{\codellama}{CodeLlama\xspace}
\newcommand{\deepseek}{DeepSeekCoder\xspace}
\newcommand{\wizard}{WizardCoder\xspace}
\newcommand{\gpt}{GPT-4o\xspace}
\newcommand{\sem}{SemCoder\xspace}
\newcommand{\starcoder}{StarCoder2\xspace}

\definecolor{problemblue}{RGB}{100,134,158}
\definecolor{idiomsgreen}{RGB}{0,162,0}
\definecolor{exercisebgblue}{rgb}{0,  .69,  .941}
\definecolor{deepgreen}{rgb}{0.0, 0.5, 0.0}
\definecolor{codegreen}{rgb}{0,0.6,0}
\definecolor{codegray}{rgb}{0.5,0.5,0.5}
\definecolor{codepurple}{rgb}{0.58,0,0.82}
\definecolor{backcolour}{rgb}{0.95,0.95,0.92}
\definecolor{redColor}{RGB}{255,0,0}
\definecolor{Gray}{gray}{0.1}
\definecolor{CadetBlue}{RGB}{243.0,42.1,134.0}

\newtcbtheorem[]{Theorem}{}{
        enhanced,
        sharp corners,
        colback=white,
        colframe=exercisebgblue
    }{thm}

\lstdefinestyle{code}{
  backgroundcolor=\color{gray!4},
  commentstyle=\color{codegray},
  keywordstyle=\color{codepurple},
  numberstyle=\tiny\color{codegray},
  stringstyle=\color{codegray},
  basicstyle=\ttfamily\footnotesize,
  breakatwhitespace=false,         
  breaklines=true,                 
  captionpos=b,                    
  keepspaces=true,                 
  numbers=left,                    
  numbersep=5pt,                  
  showspaces=false,                
  showstringspaces=false,
  showtabs=false,                  
  tabsize=2,
}
\lstdefinelanguage{test}{%
	language     = python,
	breaklines = true,backgroundcolor=\color{white},escapechar=!,rulecolor=\color{black}, breaklines=true,sensitive=true,  numbersep=5pt, xleftmargin=.015\textwidth, frame=tb,label=test
}
\lstdefinelanguage{source}{%
	language     = python,
	breaklines = true,
firstnumber=0,numberfirstline=false,columns=fullflexible,numbers=left,backgroundcolor=\color{white},
    rulecolor=\color{black}, 
    breaklines=true,sensitive=true, numbersep=5pt, xleftmargin=.015\textwidth, label=test
}
\newcommand*\Suppressnumber{%
  \lst@AddToHook{OnNewLine}{%
    \let\thelstnumber\relax%
     \advance\c@lstnumber-\@ne\relax%
    }%
}

\newcommand*\Reactivatenumber{%
  \lst@AddToHook{OnNewLine}{%
   \let\thelstnumber\origthelstnumber%
   \advance\c@lstnumber\@ne\relax}%
}

\definecolor{diffstart}{named}{codegreen}
\definecolor{diffincl}{named}{redColor}

\DeclareCaptionFormat{listing}{#3}
\graphicspath{{./figures/}}

\newcounter{NumObservations}
\stepcounter{NumObservations}
\definecolor{shadecolor}{rgb}{.9,.9,.9}

\definecolor{msftBlue}{RGB}{0,164,239}
\definecolor{msftGreen}{RGB}{127,186,0}
\definecolor{msftYello}{RGB}{255,185,0}
\usepackage{bigstrut}

\definecolor{vlcolor}{rgb}{0.9,0.1,0.1}

\begin{document}


\title[]{Narrowing the Complexity Gap in the Evaluation of Large Language Models}


\author{Yang Chen}
\affiliation{
  \institution{{\mbox{University of Illinois Urbana-Champaign}}}
  \country{USA}
}
\email{yangc9@illinois.edu}

\author{Shuyang Liu}
\affiliation{
  \institution{{\mbox{University of Illinois Urbana-Champaign}}}
  \country{USA}
}
\email{sl225@illinois.edu}

\author{Reyhaneh Jabbarvand}
\affiliation{
  \institution{{\mbox{University of Illinois Urbana-Champaign}}}
  \country{USA}
}
\email{reyhaneh@illinois.edu}

\begin{abstract}

Evaluating Large Language Models (LLMs) with respect to real-world code complexity is essential. Otherwise, there is a risk of overestimating LLMs’ programming abilities based on simplistic benchmarks, only to be disappointed when using them in real-world settings. Recently, researchers explored the construction of more realistic benchmarks by mining or augmenting open-source repositories. Such solutions are usually task-specific. Data quality control from real-world projects can also be time-consuming and error-prone. More importantly, evaluating LLMs on fixed benchmark problems is subject to data contamination and overfitting. 

We propose \approach, an automated technique to add real-world complexities to any programming benchmark. \approach leverages a multi-objective optimization to \emph{increase the complexity} of programming problems while \emph{maintaining the readability} of code similar to real-world programs. Transforming \emph{four} widely-used programming benchmarks using \approach and evaluating \emph{$13$} LLMs (including \emph{two} reasoning LLMs) on them shows a \emph{notable performance drop} across all programming tasks ($14.9\%$--$60.5\%$, avg$=35.2\%$), demonstrating LLMs' struggle under real-world complexities. The struggle persists even when LLMs are few-shot prompted or fine-tuned with examples from different versions of \approach, demonstrating the challenging nature of the problems. Finally, we show that the performance of the studied LLMs in bug repair is similar under \approach and SWE-Bench. This, along with the consistent reproduction of performance drop of all studied LLMs across four tasks under different versions of \approach, makes the technique suitable to evaluate LLMs without costly construction of real-world benchmarks.
\end{abstract}




\maketitle

\vspace{-5pt}
\section{Introduction}
\label{sec:introduction}

Large Language Models (LLMs) have demonstrated notable capabilities in various programming tasks~\cite{yang2024qwen2,roziere2023code,zhu2024deepseek,achiam2023gpt,lozhkov2024starcoder,reid2024gemini,schafer2023empirical,ibrahimzada2024repository,yang2024vert,liu2024codemind,yang2025swe}, mostly measured based on their performance on simple benchmarks. Whether their performance can generalize to real-world settings and their \emph{inherent complexities} is not rigorously explored. 
Without benchmarks reflecting the complexities of real-world programs, researchers risk overestimating LLMs' programming abilities based on simplistic benchmarks, leading to unrealistic expectations and potential disappointment in real-world settings. 

Recent efforts facilitate the evaluation of LLMs for specific programming tasks under real-world settings: SWE-Bench~\cite{jimenez2024swebench} mines GitHub repositories and artifacts to construct a benchmark of $2,294$ issue and pull request pairs from $12$ Python projects. R2E~\cite{jain2024r2e} collects methods with docstrings from GitHub repositories, improves the docstrings, and augments methods with the equivalence test harness. Using the methods, their corresponding improved docstrings, and tests---along with the repositories as context---R2E constructs a benchmark of $246$ problem specifications from $137$ Python projects. While these approaches represent a significant step toward more accurate LLM evaluation, they still face several limitations: 

\vspace{3pt}
\noindent \textbf{L1- Task-specific Evaluation.} SWE-Bench (or SWE-PolyBench~\cite{rashid2025swe}) can evaluate LLM abilities in bug-related tasks, although it is mainly used to assess program repair capabilities. SWE-Bench lacks adequate regression or reproduction tests to support other programming tasks. R2E is designed to evaluate LLM abilities in program synthesis. Due to the generation of high coverage tests for methods, R2E \emph{potentially} can be used for tasks other than program synthesis. However, supporting other programming tasks can be \emph{tricky}. For example, while R2E generates tests for the source language and claims these tests can be used in validating code translations, several research projects have shown that the automated generation of equivalent tests in the target language \emph{in the context of real-world projects} is non-trivial~\cite{pan2024lost,ibrahimzada2024repository,zhang2024scalable,yang2024vert}, requiring manual validation. New programming tasks, such as code reasoning~\cite{gu2024cruxeval,liu2024codemind,chen2024reasoning}, require extracting ground-truth input and output values from tests or collecting dynamic values based on test execution to prompt LLMs for prediction. This is again non-trivial in real-world projects due to the high degree of inter-procedural dependencies and the need for complex setups such as configuring and connecting to servers. 

\vspace{1pt}
\noindent \textbf{L2- Inefficient Quality Control.} Even for a specific task, while mining repositories 
can be automated, validating the quality of the collected data requires extensive, mostly manual effort. The original SWE-Bench dataset suffered from overly specific or unrelated tests, ambiguous issue descriptions, and reproducibility issues~\cite{swe-bench-verified}. 
A group of professionals manually reviewed and annotated a subset of $500$ problems,
resulting in SWE-Bench Verified~\cite{swe-bench-verified}. Recently, researchers identified solution leakage---the issue report or the comments explicitly solving the issue---in the SWE-Bench dataset, including the SWE-Bench Verified. Removing those cases, the reported performance of LLMs drastically dropped~\cite{aleithan2024swe}. Over time, the benchmarks can mature with fewer quality problems. However, new problems of data contamination and overfitting arise. 

\vspace{1pt}
\noindent \textbf{L3- Data Contamination.} Data contamination---the problem of training and testing data overlap---is inevitable regarding LLMs, as most models are trained on open-source code. One solution to account for data contamination is testing LLMs on GitHub projects developed after the model's training date. However, the rapid advancements of LLMs can impact the diversity of the selected benchmark, resulting in immature and not very complex repositories, hence, challenging LLMs less~\cite{concerned2024cao}. As of today, more than $90\%$ of SWE-Bench problems are dated \emph{before} the training cut-off date of most recent LLMs, indicating that updating mined real-world repositories is not practical. 

\vspace{1pt}
\noindent \textbf{L4- Overfitting.} Even if data contamination is taken into account, over time, LLM-based solutions overfit to the benchmarks, resulting in inflated performance metrics that do not generalize well to unseen patterns or contexts~\cite{kiela2023plottingprogress}. For Code LLMs, despite using real-world projects in the benchmarks, LLMs may succeed in solving them, but not the same problem outside of it. 

To overcome the issues mentioned, we introduce \approach, an automated technique for generating programming benchmarks that are closer to real-world problems in terms of complexity, without the need for mining. \approach can take any existing programming benchmark\footnote{The current implementation of \approach supports Python as the dominant programming language in major benchmarks.} and evolve its problems using a multi-objective genetic algorithm to \emph{increase their complexity} while \emph{maintaining the readability/naturalness} of the generated programs as a conflicting objective. Once evolution terminates, the optimally transformed programming problems on the Pareto front form the final version of the given benchmark, now more representative of real-world complexity. \approach is task-agnostic, overcoming \textbf{L1 limitation}. It also builds on existing mature programming benchmarks, minimizing quality control efforts and mitigating \textbf{L2 limitation}. 
\approach overcomes \textbf{L3 limitation} by upgrading programs to more complex versions with several non-trivial changes. \approach programs are also generated through the non-deterministic application of $22$ genetic operators; re-executing \approach results in different programs with different complexity patterns and embedding representation, overcoming the \textbf{L4 limitation}. 

Evaluation results show that \approach effectively enhances the complexity of original programs while preserving readability in four existing benchmarks, namely \avatar~\cite{ahmad2021avatar}, \classeval~\cite{du2023classeval}, \cruxeval~\cite{gu2024cruxeval}, and \humaneval~\cite{chen2021evaluating} (\S \ref{subsec:rq1}).  
Evaluating $13$ LLMs on \emph{four diverse} programming tasks (input prediction, output prediction, code translation, and bug repair) on transformed datasets shows a significant drop ($14.9\%$--$60.5\%$, avg$=35.2\%$) in LLMs' performance (\S \ref{subsec:rq2}). \approach can serve as a long-term evaluation benchmark: when presented with transformations of the \emph{same} or \emph{different} programs through fine-tuning or few-shot learning~\cite{brown2020language}, LLMs still struggle and demonstrate $4.7\%$--$30.3\%$ (avg$=17\%$) and $17.4\%$--$65.2\%$ (avg$=41\%$) performance drop in fine-tuned models and few-shot prompting, respectively (\S \ref{subsec:rq3}). 
Finally, \approach problems can challenge LLMs in bug repair as much as SWE-Bench, i.e., LLMs' performance on \approach indicates their performance in real-world settings without the need for new real-world benchmarks (\S \ref{subsec:rq4}).

\approach is the first technique to fully automate the transformation of code-to-code/text problems\footnote{\approach is orthogonal to research on the automated generation of program synthesis (text-to-code) problems~\cite{evoeval,chen2024ppm}.} to narrow the complexity gap between the programs in well-known benchmarks while maintaining readability. The power of \approach is in formulating the problem as a multi-objective optimization, transforming programs at several places using $22$ unique genetic operators (e.g., introducing concurrency, API dependencies, and decorators), and guiding the transformations using two novel complexity and readability metrics (\S \ref{sub:evolution}). \approach can be reconstructed without the need for mining or tracking training cut-off dates of LLMs. The engineered combination of added complexities results in programs that can challenge LLMs even under few-shot settings.  
\vspace{-8pt}
\section{Problem Statement and Challenges}
\label{sec:motivation}

\sloppy To emphasize the lack of real-world complexities in popular task-specific benchmarks (\humaneval and \classeval for code synthesis, \avatar for code translation, and \cruxeval for code reasoning), we compared their complexities with SWE-Bench\footnote{As of the time of paper submission, R2E-Eval~\cite{jain2024r2e} benchmark is not publicly available~\cite{r2e-issue}.}. We identify the following set of \emph{seven} quantifiable code complexity metrics, listed in Table~\ref{metrics-complexity}:
\textit{Cyclomatic Complexity} ($C_1$), \textit{Compound Predicates} ($C_2$), \textit{Nested Constructs} ($C_3$), \textit{Structural Complexity} ($C_4$), \textit{Third-party APIs } ($C_5$), \textit{Inter-class Dependencies} ($C_6$) and \textit{Intra-class Dependencies} ($C_7$). A \emph{higher} number for each metric indicates \emph{more} complexity. 

\begin{table}[t]
\centering
\vspace{13pt}
\caption{Code complexity metrics.}
\vspace{-8pt}
\label{metrics-complexity}
\footnotesize
\setlength{\tabcolsep}{0.5pt}
\renewcommand{\arraystretch}{0.9} 
\begin{tabular}{|c|p{3cm}|p{9cm}|} 
    \hline
    \multicolumn{1}{|c|}{\textbf{Metric}} & \multicolumn{1}{c|}{\textbf{Property}} & \multicolumn{1}{c|}{\textbf{Description}} \\
    \hline
    $C_1$ & Cyclomatic Complexity & 
    For a Control Flow Graph (CFG) of a program with $N$ nodes, $E$ edges, and $P$ connected components, $C1 = E - N + 2 * P$. \\
    \hline
    $C_2$ & Compound Predicates & 
    Total number of compound predicates (sub-predicates connected with boolean, comparison, binary, or unary operators). \\
    \hline
    $C_3$ & Nested Constrcuts & 
    Total nested levels in recursive (\texttt{for} and \texttt{while} loops) and conditional constructs. \\
    \hline
    $C_4$ & Structural Complexity & 
    Total number of complex code structures: list comprehension, dict comprehension, set comprehension, generator, lambda expression, list, thread, recursive function, and decorator. \\
    \hline
    $C_5$ & Third-party APIs & 
    Total number of third-party API calls. \\
    \hline
    $C_6$ & Inter-class Dependencies & 
    Total number of inter-class dependencies. \\
    \hline
    $C_7$ & Intra-class Dependencies & 
    Total number of intra-class dependencies. \\
    \hline
\end{tabular}
\vspace{-15pt}
\end{table}

$C_1$, $C_6$, and $C_7$ are classic metrics for measuring code complexity~\cite{mccabe1976complexity}. We also consider four LLM-specific code complexities: LLMs can struggle when the code consists of compound predicates, deeply nested constructs, and complex code structures such as list comprehension~\cite{liu2024codemind,hooda2024large,ni2024next}. For specific tasks, LLMs should know third-party APIs to avoid translation hallucination~\cite{pan2024lost} or reasoning shortcuts~\cite{li2023deceptive}. Also, while Cognitive Complexity~\cite{campbell2018cognitive} combines $C_1$ and $C_3$, we deliberately separated them into two metrics to better emphasize each complexity dimension, i.e., $C_1$ focuses on the control flow, while $C_3$ highlights nesting. Otherwise, through the evolution of the programs, $C_3$ can be dominated by $C_1$, resulting in programs with fewer nested contracts. 

\begin{figure*}[t]
\vspace{-2pt}
\includegraphics[width=0.95\textwidth]{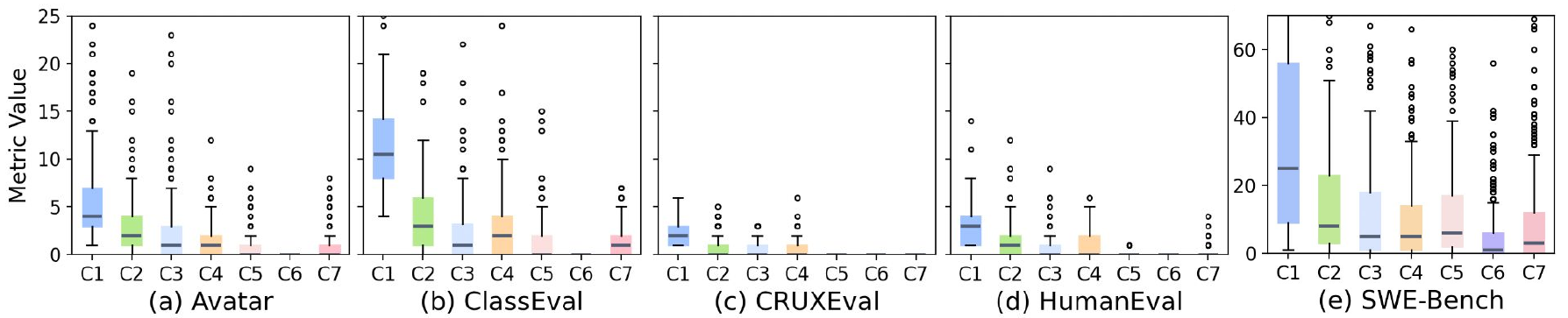}
    \vspace{-8pt}
    \caption{Code complexity of existing benchmarks}
    \vspace{-8pt}
    \label{fig:benchmark-overview}
\end{figure*}

Figure~\ref{fig:benchmark-overview} shows the distribution of complexity metric values for selected benchmarks and how they compare with a real-world benchmark such as SWE-Bench\footnote{For SWE-Bench, we measure complexity metrics for the buggy classes from files that were involved in the commit diffs rather than the entire projects.} (Figure~\ref{fig:benchmark-overview}e). It is evident that the majority of existing benchmarks do not represent real-world complexity, i.e., SWE-Bench consistently exhibits higher complexity across all seven metrics. 
Notably, all benchmarks have zero inter-class dependencies ($C_6$). \cruxeval also lacks third-party API calls ($C_5$) and intra-class dependencies ($C_7$). This shows a potential misalignment between most benchmarks and the complex nature of software in practice, resulting in likely low performance when deployed in real-world settings. \approach aims to evolve arbitrary well-crafted benchmarks for different code-to-code/text tasks into their more complex yet readable versions, enabling the evaluation of LLMs under real-world complexity. To that end, it should overcome three challenges:

\begin{wraptable}{t}{0.55\textwidth}
\centering
\scriptsize
\vspace{4pt}
\caption{Code readability metrics.}
\vspace{-8pt}
\label{metrics-readability}
    \setlength{\tabcolsep}{0.5pt}
    \renewcommand{\arraystretch}{0.8}
\begin{adjustbox}{width=0.55\textwidth,center}
    \begin{tabular}{|c|p{2.5cm}|p{3cm}|}  
    \hline
    \multicolumn{1}{|c}{\textbf{Metric}} & \multicolumn{1}{|c|}{\textbf{Property}} & \multicolumn{1}{c|}{\textbf{Description}} \\
    \hline
    $R1$ & Tokens & Total number of tokens. \\
    \hline
    $R2$ & Lines of Code & Total number of lines of code. \\
    \hline
    $R3$ & Primitive Variables & Total number of primitive type variables. \\
    \hline
    $R4$ & Compound Variables & Total number of compound variables, e.g., arrays. \\
    \hline
    $R5$ & Operators & Total number of operators. \\
    \hline
    $R6$ & Conditional Statements & Total number of \texttt{if} statements. \\
    \hline
    $R7$ & Loops & Total number of \texttt{while} and \texttt{for} loops. \\
    \hline
    $R8$ & Assignments & Total number of assignments. \\
    \hline
    $R9$ & Max Nested Loop Level & Maximum nested levels of \texttt{for} and \texttt{while} loops. \\
    \hline
    $R_{10}$ & Max Nested \texttt{if} Level & Maximum nested levels of \texttt{if} statements. \\
    \hline
    $R_{11}$ & Max Tokens per Line & Maximum number of tokens per line. \\
    \hline
    $R_{12}$ & Nested Casting Statements & Total number of statements with nested casting. \\
    \hline
    $R_{13}$ & Naturalness & Entropy of the program. \\
    \hline
    \end{tabular}
    \end{adjustbox}
    \vspace{-10pt}
\end{wraptable}

\textbf{Benchmark Saturation.} \sloppy One of the challenges in LLM evaluation is overfitting to benchmark problems. This is exacerbated, as LLMs can easily capture the patterns in the benchmarks, resulting in rapid benchmark saturation~\cite{kiela2023plottingprogress}. To address this challenge, the current version of \approach implements $22$ transformation operators (\S \ref{sec:all-operators}) and evolves programs using a non-deterministic combination of them to maximize the complexity. Even when provided with \approach transformations as in-context examples for few-shot learning, LLMs still struggle in all studied programming tasks under other \approach's transformations (\S \ref{subsec:rq3}). \approach is also fully automated and scalable (\S \ref{sec:approach}), helping users to regenerate complex transformations efficiently to account for data contamination in newer models. 

\textbf{Reusing Benchmark Artifacts.} Automated augmentation of programs with artifacts to enable multi-task evaluation, e.g., high-coverage tests with high-quality assertions, complex and representative bugs, and equivalent translations, is quite challenging for real-world programs compared to simple ones. As a result, \approach maintains the functionality of original programs through semantically equivalent transformations to re-use their artifacts. Compared to techniques introducing small semantically-equivalent perturbations to programs~\cite{li2022closer,concerned2024cao, rabin2021generalizability,pour2021search}, our transformations are more complex and need non-trivial flow- and context-sensitive program analysis (\S \ref{sec:all-operators}). Without such analyses, the genetic algorithm may result in many undesirable and useless transformations.

\textbf{Readability and Naturalness.} Higher complexity can sacrifice readability. For example, increasing nested levels of loops ten times results in complex but unreadable code, unlike what developers write. Generating readable and natural programs is specifically important in \approach, as the goal is to challenge LLMs for generalizability to real-world complexity. Without maintaining readability, one cannot conclude LLM's performance drop is solely due to increased complexity. To account for this challenge, \approach formulates the evolution as a multi-objective optimization, maximizing the complexity and maintaining readability simultaneously (\S \ref{subsec:GA}). Once \approach terminates, the solutions in the Pareto frontier are complex and readable, similar to real-world programming problems (\S \ref{subsec:rq1}).
\begin{wrapfigure}{r}{0.55\columnwidth}
\centering
\vspace{-8pt}
\noindent\begin{minipage}{.95\linewidth}{
\begin{lstlisting}[language = python,basicstyle=\small, style=code,numbersep=5pt,showlines=true,columns=fullflexible,escapechar=|]
|\addgreen|+ import queue, threading
|\addgreen|+ from data_transformer_class import 
|\addgreen|  data_transformer
|\addgreen|+ from scipy.stats import ttest_ind
|\addgreen|+ def my_decorator(func):
|\addgreen|+    def dec_result(*args, **kwargs):
|\addgreen|+        processed_result = func(*args, **kwargs)
|\addgreen|+        return processed_result
|\addgreen|+    return dec_result
|\addgreen|+ @my_decorator
|\addred|- def f(text, char):
|\addgreen|+ def f(text_data, char):
|\addred|-    if text:
|\addgreen|+    if text_data:
|\addred|-         text = text.removeprefix(char)
|\addgreen|+         text_data = text_data.removeprefix(char)
|\addred|-         text = text.removeprefix(text[-1])
|\addgreen|+         text_data = text_data.removeprefix(
|\addgreen|          text_data[-1])
|\addred|-         text = text[:-1] + text[-1].capitalize()
|\addgreen|+         modified_text = text_data[:-1]
|\addgreen|+         add_data = text_data[-1]
|\addgreen|+         result_queue = queue.Queue()
|\addgreen|+         def processing_thread(queue):
|\addgreen|+             result = data_transformer(
|\addgreen|              modified_text, add_data)
|\addgreen|+             queue.put(result)
|\addgreen|+         worker_thread = threading.Thread(target = 
|\addgreen|               processing_thread, args=(
|\addgreen|               result_queue,))
|\addgreen|+         worker_thread.start()
|\addgreen|+         worker_thread.join()
|\addgreen|+         func_result = result_queue.get()
|\addgreen|+         text_data = func_result
|\addgreen|+     ttest_ind([31, 91, 49], [39, 26, 7])
|\addgreen|+     return text_data
# In data_transformer_class.py:
|\addgreen|+ from sklearn.utils import shuffle
|\addgreen|+ def data_transformer(data, add_data):
|\addgreen|+    shuffle([71, 26, 77])
|\addgreen|+    return data + add_data.capitalize()
\end{lstlisting}}
\end{minipage}
\vspace{-8pt}
\caption{\approach transformation of CRUXEval-409}
\label{fig:code-example1}
\end{wrapfigure}

Based on a comprehensive study of existing readability studies~\cite{Buse2008metric,Scalabrino2017Automatically, Hindle2012naturalness, posnett2011simpler} with Python-specific considerations, we identify \emph{$13$} quantifiable readability metrics (Table~\ref{metrics-readability}), measuring the number of \emph{tokens} ($R_1$), \emph{lines of code} ($R_2$), \emph{primitive type variables} ($R_3$), \emph{compound variables} ($R_4$), \emph{operators} ($R_5$), \emph{conditional statements} ($R_6$), \emph{loop statements} ($R_7$), \emph{assignment statements} ($R_8$), \emph{maximum nested levels} in loops ($R_9$) and conditional statements ($R_{10}$), \emph{maximum tokens per each line} ($R_{11}$), and \emph{nested casting statements} ($R_{12}$). We also measure the \emph{naturalness} through program entropy ($R_{13}$)~\cite{adler2021improving,posnett2011simpler}. A \emph{higher} number for each metric indicates \emph{less} readability. \approach uses readability and complexity metrics (Table~\ref{metrics-complexity}) in the objective function to guide the benchmark generation (\S \ref{sub:fitness}).

Figure ~\ref{fig:code-example1} illustrates \approach transformation of \texttt{\small sample_409} program in \cruxeval. The highlighted \textcolor{textred}{red} lines represent the original program, and \textcolor{textgreen}{green} lines are added by \approach. The transformations, which we later explain in detail (\S \ref{sec:all-operators}), include (1) renaming variables, (2) extracting binary operations as a new function (intra-class dependency), (3) moving it into a new module to create inter-class dependency, (4) generating new assignments, (5) introducing multi-threading, (6) introducing a decorator, and (7) adding third-party API calls, i.e., \texttt{scipy} and \texttt{scikit-learn}. The transformation is functionally equivalent to the original program but \emph{20 times} more complex (\S \ref{sub:fitness}). 

\begin{wrapfigure}{r}{0.55\columnwidth}
\centering
\noindent\begin{minipage}{.95\linewidth}{
\begin{lstlisting}[language = python,basicstyle=\small, style=code,numbersep=5pt,showlines=true,columns=fullflexible,escapechar=|]
import threading
from .standard_profile import ThreadingXMLRPCServer
def __init__(self,...):
    if self._callable:
|\addblue|            # Feature: Threads
            self._thread = threading.Thread(target=
                 self._serve_forever)
            self._thread.daemon = True
|\addblue|            # Feature: Inter-Class dependencies       
            self.client = ThreadingXMLRPCServer()
def start(self):
    if self._callable:
        self._is_running = True
|\addblue|        # Feature: Intra-Class dependencies
        self._run_client()
def _run_client(self):
    if self._callable:
        self._thread.start()
|\addblue|# Feature: Decorators
@property
def is_registered(self):
    return self._is_registered
\end{lstlisting}}
\end{minipage}
\vspace{-12pt}
\caption{Real-world code example from Astropy~\cite{astropy}}
\vspace{-10pt}
\label{fig:realworld-code-example}
\end{wrapfigure}
The readability of \approach transformation is similar to a real-world example in Figure~\ref{fig:realworld-code-example} (from project Astropy~\cite{astropy}), with PyLint score $9.62$ and $9.45$, respectively (out of $10$), which shows adding complexity by \approach does not result in an unreadable spaghetti code. 
The two programs share similar non-arbitrary features such as decorators, multi-threading, and inter/intra-class dependencies (highlighted in \textcolor{textblue}{blue}), showing that \approach operators can reproduce real-world patterns and represent real-world programs.

\section{\approach}
\label{sec:approach}

\approach takes programs from an existing benchmark as input and iteratively evolves them to (1) increase complexity while (2) maintaining readability.
In this section, we first describe the formulation of the benchmark generation problem as a multi-objective optimization by (\S \ref{subsec:GA}), and explain the details of fitness evaluation (\S \ref{sub:fitness}) and evolution process (\S \ref{sub:evolution}). 

\subsection{Genetic Algorithm}
\label{subsec:GA}

\begin{wrapfigure}{r}{0.55\columnwidth}
  \vspace{-15pt} 
  \includegraphics[width=0.55\columnwidth]{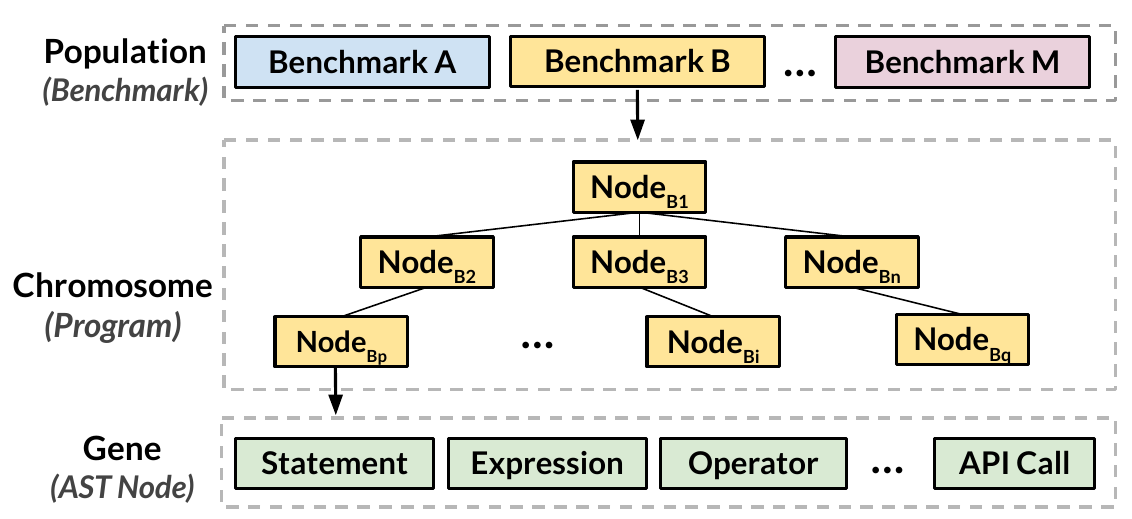}
  \vspace{-20pt}
  \caption{Genetic makeup in \approach}
  \label{fig:approach}
\end{wrapfigure}
Figure~\ref{fig:approach} illustrates the genetic representation of the evolutionary benchmark generation. Each program in the benchmark is a chromosome, and a chromosome consists of genes, which are Abstract Syntax Tree (AST) nodes of the program. The sub-genes could be statements, expressions, operators, method calls, etc. \approach (Algorithm~\ref{alg:genetic-algorithm}) begins with an initial population of simple programs from an existing benchmark $B$, and evolves programs in parallel. At each iteration (Lines 5--12), it selects the best chromosomes in the current population (Line 6), i.e., those with the highest complexity and readability (\S \ref{sub:evolution}). It then modifies them with applicable transformation operators to generate new offspring (Line 7). 

At this stage, \approach performs two validity checks (Line 8) and discards transformations that are (1) not functionally equivalent to the original program or (2) have a readability below the threshold $RT$ or lower than the original program (\S \ref{sub:fitness}). At the end of each iteration, \approach adds the new offspring to the population (Line 9).
The evolution of each program in the population continues until one of the following criteria is met: (1) there exists a program in the population that reaches the complexity threshold $CT$ (Lines 10--11) or (2) \approach reaches the given time budget. Once evolution terminates, \approach selects the candidate with the highest complexity among those on the Pareto front (Line 13) and returns the transformed benchmark (\S \ref{sub:evolution}).

\vspace{20pt}
\subsection{Fitness Evaluation}
\label{sub:fitness}

\approach rewards transformations according to their \emph{complexity} and \emph{readability}. The higher the complexity of a transformation, the more likely it is to evolve in the next iteration and be selected as a final program. Increasing the complexity, however, can reduce readability; e.g., nesting constructs beyond a threshold makes the code less readable or is uncommon in programs written by developers. To control for readability and complexity of the transformations, \approach measures them \emph{relative} to problems in real-world programming benchmarks. 

Considering a set of complexity metrics $\phi=\{C_1, \ldots, C_n \}$ (Table \ref{metrics-complexity}) and readability metrics $\psi=\{R_1, \ldots, R_m \}$ (Table \ref{metrics-readability}), \approach computes the average of each metric $C_i$ or $R_i$ across all buggy classes in SWE-Bench\footnote{R2E dataset is not publicly available.}. Using the calculated values, it constructs representative complexity and readability thresholds, $\phi (CT)=\{CT_{1}, \ldots, CT_{n}\}$ and $\psi (RT)=\{RT_{1}, \ldots, RT_{m}\}$, respectively. 
After calculation of representative thresholds for complexity and readability, for each program $P$ with complexity $\phi (P)=\{C_{1}, \ldots, C_{n}\}$ and readability $\psi (P)=\{R_{1}, \ldots, R_{m}\}$, \approach calculates \emph{Relative Complexity (RC)} and \emph{Relative Readability (RR)} as below:

\vspace{-5pt}
{
\small
\begin{equation} 
\label{eq:relative-comp}
    \text{RC}(P) = \frac{1}{n} \sum_{i=1}^n RC_i, \quad
    RC_i = 
    \begin{cases}
        \frac{C_i}{CT_i} & \text{if } C_i < CT_i \\
        1 & \text{if } C_i \geq CT_i 
    \end{cases}
\end{equation}
}
\vspace{-6pt}
{
\small
\begin{equation}
\label{eq:relative-read}
    \text{RR}(P) = \frac{1}{m} \sum_{i=1}^m RR_i, \quad
    RR_i = 
    \begin{cases} 
        1 - \frac{R_i}{RT_i} & \text{if } R_i < RT_i \\
        0 & \text{if } R_i \geq RT_i 
    \end{cases}
\end{equation}
}
\begin{wrapfigure}{r}{0.7\textwidth}
\begin{algorithm}[H] 
\footnotesize
\caption{Evolutionary Benchmark Generation}
\label{alg:genetic-algorithm}
\KwIn{Original Benchmark $B$; Transformation Operators $TO$; Complexity Threshold $CT$; Readability Threshold $RT$; Breed Size $k$; $Budget$;}

\KwOutput{Transformed Benchmark $TB$}

$TB \gets B$

\ForEach{$p_i \in TB$}
{

$Population_i \gets p_i$

$complexity_{reached} \gets TRUE$
 
 \While{$ complexity_{reached} \,\&\, Budget \ge 0$} 
{
    $Selection_i \gets \text{SelectToEvolve}(Population_i, CT, RT, k)$
    
    $Offspring_i \gets \text{Transform}(Selection_i, TO)$

    $ValidOffspring_i \gets \text{ValidateOffspring}(Offspring_i, Population_i, RT)$

    $Population_i \gets Offspring_i \cup Population_i$

    \If{$ \exists \, pp_{ij} \in Population_i$ where $C_{pp_{ij}} \ge CT$}{$complexity_{reached} \gets FALSE$}

    $Budget \mathrel{-}= 1$
}

$p_i \gets selectTopCandidate(Population_i,CT,RT)$

}

\KwRet $TB$
\end{algorithm}
\vspace{-10pt}
\end{wrapfigure}


The $RC$ and $RR$ values belong to $[0, 1]$. For complexity metrics $C_i$s, the higher and closer values to corresponding $CT_i$s result in higher $RC$. For readability metrics $R_i$s, the lower and farther values to corresponding $RT_i$s result in higher $RR$. \approach uses $RC$ and $RR$ values for (1) fitness evaluation (transformations with higher RC and RR values are more likely selected to breed the next generation), (2) validating generated offspring (\approach discards an offspring $P$ when $\exists  \, R_i, \, RR_i=0$), and (3) terminating the evolution (\approach terminates the evolution of a program once for at least one population individual $P$, $RC(P)=1$, i.e., $\forall i \in n, \, C_i \ge CT_i$).

\vspace{-8pt}
\subsection{Evolution}
\label{sub:evolution}

\approach evolves population using two sets of genetic operators at each iteration: the selection operator to select higher quality chromosomes from the population to breed the next generation or final selection of elite chromosomes (\S \ref{selection-details}), and the transformation operator to modify genes and sub-genes of selected chromosomes to produce offspring (\S \ref{mutation-details}). 

\subsubsection{Chromosome Selection}
\label{selection-details}

\approach implements a roulette wheel selection, i.e., fitness-proportionate selection strategy, to breed offspring at each iteration (function \texttt{\small SelectToEvolve} in Line 6 of Algorithm~\ref{alg:genetic-algorithm}). That is, the likelihood of selecting a chromosome is proportional to its fitness value. At the beginning of each iteration, for each population individual---transformations of a given program in the original benchmark---\approach calculates \emph{Relative Complexity} ($RC$) and \emph{Relative Readability} ($RR$) values. 

\begin{wrapfigure}{r}{0.5\columnwidth}
  \vspace{-15pt} 
  \includegraphics[width=0.5\columnwidth]{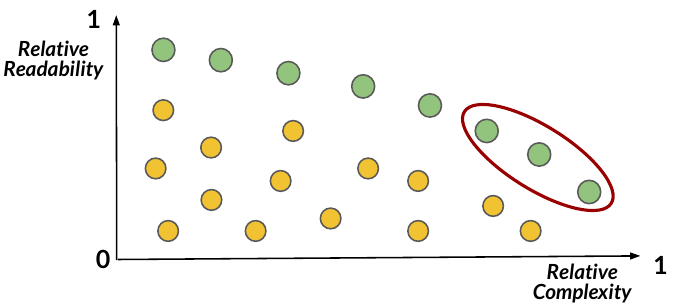}
\vspace{-25pt}
\caption{Selection from Pareto Front}
\label{fig:front}
\end{wrapfigure}

Next, \approach constructs the Pareto front following non-dominated sorting NSGA-II~\cite{deb2000fast} and selects the top $k\%$ most complex transformations to breed the next generation. 
If $k\%$ (breed size) of the population is less than one, it selects the top individual. Figure~\ref{fig:front} visualizes the chromosome selection in \approach. Transformations that appear on the Pareto front are green, and the rest of the sub-optimal transformations are yellow. \approach sorts Pareto-optimal transformations based on $RC$ and selects top $k\%$. 
Once the evolution terminates, \approach follows a similar strategy to select the final transformation for the benchmark (function \texttt{\small SelectTopCandidate} in Line 13 of Algorithm~\ref{alg:genetic-algorithm}). The only exception is that at the final selection, \approach selects the one with the highest $RC$ to appear in the benchmark.  

\begin{wrapfigure}{r}{0.62\textwidth}
\vspace{-12pt}
\begin{algorithm}[H]
\footnotesize
\caption{Program Transformation and Validation}
\label{alg:mutation}
\KwIn{Selected Population $Selection$; Transformation Operators $TO$; Readability Threshold $RT$}

\KwOutput{Generated Offspring $Offspring$}

$Offspring \gets \emptyset $

\ForEach{$P \in Selection$}
{

    
    
    $ApplicableOps \gets GetApplicableOperators(P, TO)$
    
    $AppliedOps \gets GetAppliedOperators(P)$
    
    $TransLocs \gets GetTransformedLocations(P, AppliedOps)$
    
    \ForEach{$operator \in ApplicableOps$}
    {
    
        $ApplicableLocs \gets GetApplicableLocations(P, operator)$
    
        $ValidLocs \gets ApplicableLocs \setminus TransLocs$
    
        \If{$ValidLocs = \emptyset $}{continue}
    
        $Location \gets RandomSelection(ValidLocs)$
        
        $P_{operator} \gets ApplyOperator(P, operator, Location)$

        $P_{operator} \gets EnhanceNaming(P_{operator})$
    
        \If{$LowReadability(P_{operator}, RT)$}{continue}

        \If{$LowPyLintScore(P_{operator, P})$}{continue}
    
        \If {$TestFailure(P_{operator})$}{continue}
            
        $Offspring \gets Offspring \cup P_{operator}$
        
    }
}
\KwRet $Offspring$
\end{algorithm}
\vspace{-3pt}
\end{wrapfigure}

\subsubsection{Chromosome Manipulation}
\label{mutation-details}

Once \approach selects high-quality transformations from the existing population, it uses them to breed the next generation, i.e., transformations that are more complex, yet readable. Recall that \approach evolves the programs separately; hence, the population at each iteration belongs to the current transformations of a single program. The algorithm~\ref{alg:mutation} shows the transformation and validation process of individuals in a selected population. For each program $P$ in the selection, \approach gets three lists (Lines 3--5): (1) applicable operators $ApplicableOps$, (2) applied operators $AppliedOps$ that have evolved the original program until current iterations, and (3) location of transformations $TransLocs$ corresponding to each previously applied operator. It then iterates over $ApplicableOps$ and transforms $P$ only if the same operator \emph{has not been applied} to \emph{same location before} (Lines 7--11). The rationale here is to prevent \approach favoring a specific operator, ensuring diversity of transformations. When a transformation operator is applicable to multiple locations, \approach randomly selects one to proceed. This non-determinism results in a diverse set of Pareto-optimal transformations, which can alleviate the threat of overfitting (\S \ref{subsec:rq3}). 

\approach modifies variables and methods with syntactic names. 
To prevent memorization of synthetic names by LLMs and enhance the naturalness of transformations, \approach prompts an LLM to replace identifier names with natural alternatives (Line 13). After new offspring individuals are generated, \approach discards those whose (1) readability degrades below the readability threshold $RT$ (Lines 14--15), (2) their Pylint~\cite{pylint} score\footnote{PyLint scores Python files on a scale of $0$ to $10$ based on how Pythonic the code is. A value closer to $10$ indicates the code is more Pythonic.} is lower than that of the original program (Lines 16--17), or (3) fail test execution (Lines 18--19). The first two are required to prevent population bloating~\cite{doerr2017bounding} while maintaining good readability, and the latter is to ensure that transformations are functionally equivalent to the original program. Validated transformations will be selected as final offspring to be added to the population (Lines 20--21).

\vspace{-8pt}
\section{Transformation Operators}
\label{sec:all-operators}

We analyzed the code features of the most popular Python projects on GitHub to design transformation operators. We selected the top $100$ projects from the Python Package Index (PyPI)~\cite{pypi} to start and identified their corresponding GitHub repositories. Excluding the test code provided us 2,692 Python files. The authors manually examined the code styles over the course of one month to identify advanced code features. Based on the insights from the manual investigation, we developed $22$ semantic-preserving transformation operators across three categories. Table~\ref{operators} lists these operators and their brief descriptions. 

\begin{table}[t]
\centering
\caption{\approach transformation operators.}
\vspace{-10pt}
\label{operators}
\resizebox{\textwidth}{!}{
\begin{tabular}{c|c|l|c|c|c}
\hline
\hline

\multicolumn{1}{c|}{Type}
& \multicolumn{1}{c|}{ID}
& \multicolumn{1}{c|}{Transformation Operator}   
& \multicolumn{1}{c|}{Description}
& \multicolumn{1}{c|}{Analysis Sensitivity}
& \multicolumn{1}{c}{Closest Prior Analog}
\\ 
\hline

\multirow{13}{*}{\shortstack{Code\\Structures}}  
& $S_1$ & \texttt{AddNestedFor} & Add another nested \texttt{for} to an existing \texttt{for} loop & Flow & No
\\
\cline{2-6}

& $S_2$ &\texttt{AddNestedIf} & Add another nested \texttt{if} to an existing \texttt{if} statement & Flow & No\\
\cline{2-6}

& $S_3$ & \texttt{AddNestedWhile} & Add another nested \texttt{while} to an existing \texttt{while} loop & Flow & No\\
\cline{2-6}

& $S_4$ & \texttt{AddThread} & Introduce a \texttt{thread} integrated with \texttt{queue} & Flow \& Context & No
\\
\cline{2-6}

& $S_5$ & \texttt{AddTryExcept} & Add a \texttt{try-except} handler inside existing functions & Flow & No 
\\
\cline{2-6}

& $S_6$ & \texttt{CreateFunction} & Transform existing statements into new functions & Flow \& Context & a simple analog~\cite{li2022closer}
\\
\cline{2-6}

& $S_7$ &\texttt{CreateModuleDependencies} & Move  functions into another Python file & Flow \& Context & No\\
\cline{2-6}

& $S_8$ &\texttt{IntroduceDecorator} 
& Introduce a decorator & Flow \& Context & No\\
\cline{2-6}

& $S_9$ & \texttt{ReplaceNumpy} & Replace applicable built-in calculations with \texttt{Numpy} & Flow \& Context & No \\
\cline{2-6}

& $S_{10}$ &\texttt{TransformAugAssignment} & Transform augment assignment to a normal assignment & Flow & a simple analog~\cite{li2022closer}
\\
\cline{2-6}

& $S_{11}$ &\texttt{TransformLoopToRecursion} & Transform existing \texttt{for} loop into recursive functions & Flow \& Context & No\\
\cline{2-6}

& $S_{12}$ &\texttt{TransformPrimToCompound} & Transform variables of primitive types into compound types, e.g., lists &Flow & No\\
\cline{1-6}

\multirow{8}{*}{API Calls}
& $A_1$ &\texttt{AddBase64} 
& Introduce API calls from \texttt{base64} library & Context & No
\\ 
\cline{2-6}

& $A_2$ &\texttt{AddCrypto} 
&  Introduce API calls from \texttt{cryptography} library& Context & No
\\
\cline{2-6}

& $A_3$ &\texttt{AddDatetime} 
&  Introduce API calls from \texttt{datetime} library& Context & No
\\
\cline{2-6}

& $A_4$ &\texttt{AddDateutil} 
&  Introduce API calls from \texttt{dateutil} library& Context & No
\\
\cline{2-6}

& $A_5$ &\texttt{AddHttp} 
&  Introduce \texttt{http} connections & Context & No
\\
\cline{2-6}

& $A_6$ &\texttt{AddScipy} 
&  Introduce API calls from \texttt{scipy} library & Context& No
\\
\cline{2-6}

& $A_7$ &\texttt{AddSklearn} 
&  Introduce API calls from \texttt{sklearn} library & Context& No
\\
\cline{2-6}

& $A_8$ &\texttt{AddTime} & Introduce API calls from \texttt{time} library & Context& No
\\
\cline{1-6}

\multirow{2}{*}{Renaming} 
& $N_1$ &\texttt{RenameVariable} 
&  Rename existing variables & Flow & ~\cite{li2022closer,concerned2024cao, rabin2021generalizability, pour2021search, chakraborty2022natgen}
\\
\cline{2-6}

& $N_2$ &\texttt{RenameFunction} 
&  Rename existing functions & Flow \& Context & ~\cite{li2022closer,concerned2024cao, pour2021search}
\\
\hline

\hline
\end{tabular}
\label{operators}
}
\vspace{-15pt}
\end{table}

\approach traverses and modifies the Abstract Syntax Tree (AST) to apply transformations (\texttt{\small ast.NodeVisitor} for traversing AST and finding the applicable locations, and \texttt{\small ast.NodeTransformer} for updating the AST). To ensure the robustness of transformations, \approach implements the operators using flow- and context-sensitive program analysis. This requires \approach to override \texttt{\small NodeVisitor} and \texttt{\small NodeTransformer} classes to consider (1) statement orders (for flow-sensitivity) and (2) caller-callee relationship (for context-sensitivity). 
\approach differs from automated refactoring techniques~\cite{zhang2022making,ouni2016multi,zhang2024refactoring,zhang2024automated} since they aim to simplify the code structure or improve readability. In contrast, it increases complexity and results in slightly less readability than the original code (although original readability decreases, \approach programs' readability never goes below the average of real-world projects). 

The majority of \approach operators are novel ($18$ out of $22$), introducing unique aspects to the code during transformations, e.g., concurrency, decorators, and API dependencies. The other four operators may have superficial analogs in prior techniques that modify programs with semantic-preserving changes, but \approach designs more unique and advanced operators (last column of Table~\ref{operators}). For example, \citeauthor{li2022closer}~\cite{li2022closer} offers a transformation similar to $S_6$ (\texttt{\small CreateFunction}), to move assignments on the right-hand side (RHS) to a new function. They only support cases where the left-hand side (LHS) is a single variable and the RHS consists of primitive variables. $S_6$ extends to more complex scenarios where the RHS includes constants or compound types (e.g., list elements). As we will later show (\S \ref{subsubsec:embedding}), even without notable embedding distribution shift through transformations (the goal of prior research), \approach results in a significant performance drop, highlighting the importance of considering complexity in the design of operators rather than aiming for semantic equivalence. The contribution of overlapping operators to complexity is less than \approach's operators (\S \ref{subsec:rq1}), demonstrating the necessity of designing new operators.

\vspace{-5pt}
\section{Evaluation}
\label{sec:evaluation}

For a comprehensive evaluation of \approach, we investigate the following research questions:

\vspace{2pt}
\noindent \textbf{RQ1: Properties of Transformations.} To what extent can \approach improve the complexity of existing benchmarks? What are the readability properties of the \approach's transformations? 

\vspace{2pt}
\noindent \textbf{RQ2: Effectiveness and Analysis of Failure.} To what extent can \approach challenge the generalizability of LLMs to real-world complexity? What factors result in LLMs' failure to generalize to real-world complexity?

\vspace{2pt}
\noindent \textbf{RQ3: Efficacy.} Can LLMs bypass \approach transformations in future through few-shot learning?

\vspace{2pt}
\noindent \textbf{RQ4: Real-world Representativeness.} To what extent is LLMs' performance on \approach close to that of similar real-world programming problems?

\vspace{-5pt}
\subsection{Experiment Setup}

\begin{figure*}[t]
    \includegraphics[width=0.97\textwidth]{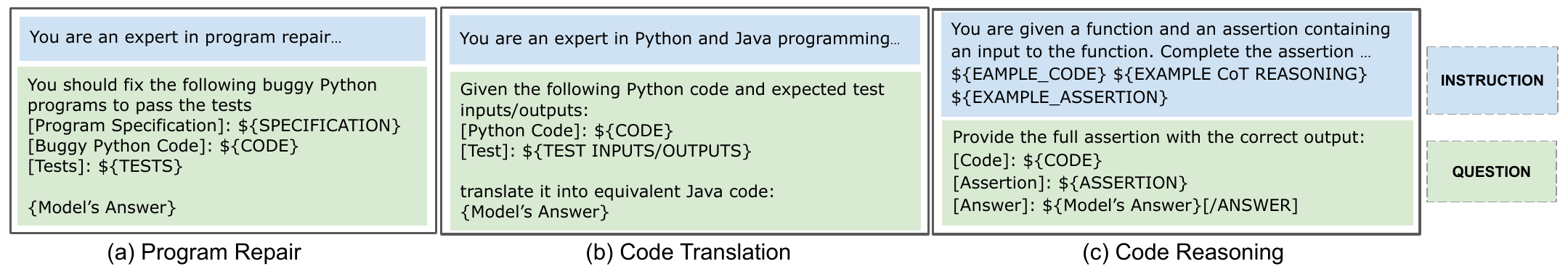}
    \vspace{-8pt}
    \caption{Prompt templates used in \approach}
    \vspace{-5pt}
    \label{fig-templates}
\end{figure*}

\approach uses Python AST library~\cite{python-ast} for static analysis and refactoring, and py2cfg~\cite{py2cfg} for control flow analysis. 
To avoid population bloating and ensure proper evolution of offspring, we set the breed size ($k$ in Algorithm~\ref{alg:genetic-algorithm}) to $20\%$. This is specifically important for our experiments, given the large number of transformation operators that generate thousands of offspring in each iteration\footnote{A smaller breed size prevents the population from being overwhelmed by excessive offspring. At the same time, it maintains diversity, since only the most promising individuals breed while weaker ones still occasionally survive through mutation~\cite{back1997handbook}.}. Following the configurations of search-based software engineering techniques, we allowed \approach to evolve transformations for one hour ($Budget$ in Algorithm~\ref{alg:genetic-algorithm}).

\vspace{-5pt}
\subsubsection{Subject LLMs} 
We selected \emph{$11$} pre-trained or instruction-tuned models of different sizes across diverse family of models: \codellama(13B-Base, 13B-Instruct, and 34B-Instruct)~\cite{roziere2023code}, \deepseek(6.7B-Base, 6.7B-Instruct, and 33B-Instruct)~\cite{bi2024deepseek,guo2024deepseek,zhu2024deepseek}, \sem  ~\cite{ding2024semcoder}, \starcoder-15b~\cite{lozhkov2024starcoder}, \wizard(15B-V1.0 and 33B-V1.1)~\cite{luo2023wizardcoder,xu2023wizardlm}, and \gpt~\cite{gpt4}. We also evaluated two reasoning models, o4-mini~\cite{o4-mini} and DeepSeekCoder-R1~\cite{guo2025deepseek}, from the same model families. We used Gemini-1.5 Pro for naturalizing variable identifiers (\S \ref{mutation-details}). To avoid any bias, we did not include any of the Gemini models among the subject LLMs. 
We set the experimental temperature to \textit{zero} to ensure the reproducibility of the results~\footnote{This applied to all $11$ pre-trained models, except the two reasoning models, which do not allow temperature changes.}. For other parameters, we used the default settings. To maximize model performance, we load the models with bfloat16 precision on $A100$ GPUs, preventing accuracy loss following the best prompting practices. Our pipeline automatically checks LLM responses to ensure they are valid and parsable, and removes such cases. That said, we observed \emph{zero} of such cases, mitigating the threat of prompt misalignment. 

\vspace{-3pt}
\subsubsection{Tasks and Subject Benchmarks} We evaluated LLMs across four code-to-code/text tasks: \emph{program repair}, \textit{code translation} (Python to Java), and \textit{code reasoning} (input prediction and output prediction). Figure~\ref{fig-templates}a shows the prompt template used for program repair, which requires program specification (to introduce expected behavior), a set of passing and failing tests, and buggy code. As a result, we chose \humaneval ($164$ programs) and \classeval ($100$ programs), two prominent function- and class-level program synthesis datasets that come with natural language specifications. 

\approach implements \emph{five} semantic-altering operators to change arithmetic/comparison/logic operators, reverse boolean value, and change variable types\footnote{We did not use HumanEvalPack~\cite{muennighoff2023octopack}, a \humaneval version with simple injected bugs, in the experiments for two reasons: First, we wanted to create similar bugs concerning complexity in both \humaneval and \classeval to enable cross analysis. Also, the injected bugs were simple, mostly changing one code location.}. It then uses them to generate higher-order mutants~\cite{jia2009higher} (max order of three) of original and transformed programs. To ensure the bugs are identical in both versions, \approach selects applicable locations from the statements shared between the original and transformed code. 

For code translation, we used the \avatar ($250$ programs) and followed best practices~\cite{pan2024lost,yang2024exploring} to prompt LLMs, e.g., including multiple tests in the prompt to help correct translation (Figure~\ref{fig-templates}b). For code reasoning, we sampled $200$ programs from the \cruxeval and used their prompt template (Figure~\ref{fig-templates}c) for input and output prediction sub-tasks. 

\approach relies on test execution to verify if transformations preserve the semantics of the original programs. To that end, the quality of the test suites is critically important. The original datasets already have very high line coverage ($90.8\%$–$99.5\%$), showing that their test suites can comprehensively exercise the programs. Importantly, \approach programs maintain comparable coverage ($93.8\%$-$99.4\%$) passing all the tests. This is because, although they are syntactically more complex with more lines, they preserve the semantics; hence, the same test will also cover the added statements. 

\vspace{-5pt}
\subsection{RQ1: Properties of Transformations}
\label{subsec:rq1}

\begin{wraptable}{r}{0.5\textwidth}
\centering
\footnotesize
\vspace{-10pt}
\caption{Relative complexity and readability of benchmarks before and after transformation. {$\Delta = (Aft - Bef)/Bef$}. }
\vspace{-10pt}
\begin{adjustbox}{width=0.45\textwidth,center}
\begin{tabular}{|l|ccc|ccc|}
\hline
\multirow{2}{*}{Benchmark} & \multicolumn{3}{c|}{Relative Complexity} & \multicolumn{3}{c|}{Relative Readability} \\
\cline{2-7}
 & Bef. & Aft. & $\Delta$ & Bef. & Aft. & $\Delta$  \\
 \hline
 
Avatar & 0.09 & 0.24 & 167\% & 0.65 & 0.55 & -15\%  \\
ClassEval & 0.16 & 0.28 & 75\% & 0.67 & 0.59 & -12\%  \\
CRUXEval & 0.02 & 0.15 & 650\%  & 0.84 & 0.75 & -11\%  \\
HumanEval & 0.04 & 0.16 & 300\%  & 0.77 & 0.69 & -10\%  \\
\hline
Average & 0.08 & 0.21 & 298\%  & 0.73 & 0.65 & -12\%  \\
\hline

\end{tabular}
\end{adjustbox}
\label{tab:metrics-comparison}
\vspace{-8pt}
\end{wraptable}

\begin{figure*}[t]
    \centering
    \includegraphics[width=0.98\textwidth]{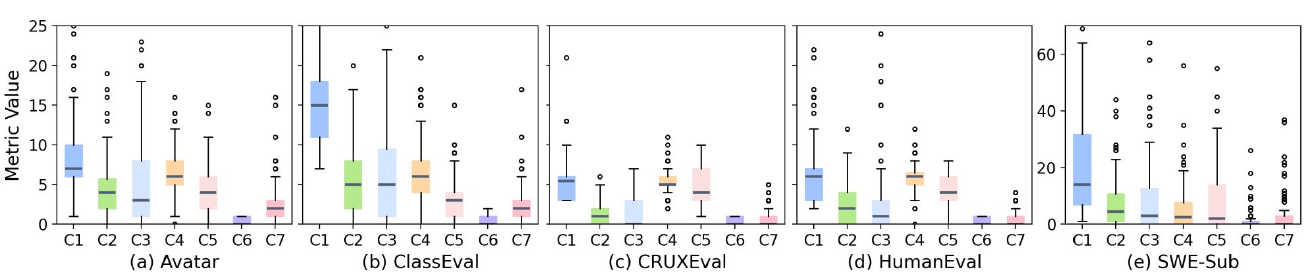}
    \vspace{-8pt}
    \caption{Code complexity of transformed benchmarks. $C_i$s are complexity metrics from Table~\ref{metrics-complexity}}
    \label{fig:after-transformation}
\end{figure*}

Figures~\ref{fig:after-transformation}a--\ref{fig:after-transformation}d illustrate the complexity of transformations for each benchmark. Compared to the programs before transformations (Figure~\ref{fig:benchmark-overview}a--Figure~\ref{fig:benchmark-overview}d), \textbf{we can observe a notable increase in each complexity metric value for \approach programs}. Notably, the original benchmarks had zero inter-class dependencies ($ C_6$). \approach increases this metric value to an average of $0.32$ across all transformations. 

Table~\ref{tab:metrics-comparison} shows the overall change in relative complexity (Equation~\ref{eq:relative-comp}) and readability (Equation~\ref{eq:relative-read}) before ($Bef.$) and after ($Aft.$) transformations. Although the transformations are far from real-world complexity in terms of metrics, i.e., average $RC = 0.21$ while the max value could be $1$, we see $298\%$ increase in the relative complexity, ranging from $75\%$ to $650\%$. This relative complexity is achieved through a one-hour time budget of \approach; hence, users of \approach can increase the budget to generate more complex transformations. As we will show, \textbf{even with an average $RC$ of $0.21$, \approach significantly challenges the generalizability of LLMs}. The relative readability declined with an average of $12\%$.
A readability drop does not imply a lack of readability, since the original programs are very simple, and a relative readability close to $0$ indicates a level similar to that of real-world programs. 
The average PyLint score of the original programs is $9.75$ (median=$10$), while the average score of \approach transformations is $9.87$ (median=$10$). These results confirm that \textbf{\approach transformations are all Pythonic}, i.e., they adhere to Python coding standards and best practices.

\begin{figure}[H]
    \centering
    \vspace{-8pt}
    \includegraphics[width=0.98\textwidth]{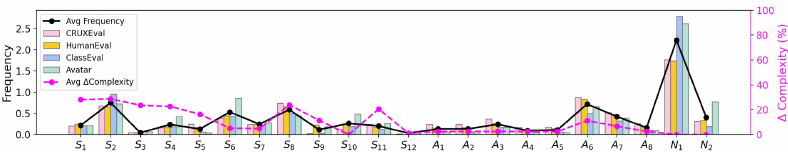}
    \vspace{-15pt}
    \caption{Frequency of operators and their contributions to overall complexity. Operator IDs are from Table~\ref{operators}}
    \label{fig:operator-frequency}
\end{figure}

Next, we investigated the frequency of each operator across the transformations and its contribution to complexity. The frequency of an operator is the total number of its occurrences divided by the total number of transformations. To estimate operators' contribution to complexity, we also applied each operator independently to all applicable programs and calculated the resulting increase in complexity. As shown in Figure~\ref{fig:operator-frequency}, \textbf{our operators are all applied to many programs in the subject benchmarks during transformations}. Spearman's rank correlation~\cite{sedgwick2014spearman} indicates no significant correlation between frequency and complexity ($\rho=0.079,$ p-value$=0.73$), suggesting that operators applied more frequently are not necessarily those introducing greater complexity. This demonstrates a well-designed fitness function that allows program complexity to grow in different complexity directions without starving some operators due to a lack of complexity. Code structure operators ($S_i$s) contribute more to complexity, especially those that introduce nested constructs, threads, and decorators.

\subsection{RQ2: Effectiveness and Analysis of Failure}
\label{subsec:rq2}

\begin{table}[t]
\centering
\setlength{\tabcolsep}{1pt}
\footnotesize
\caption{LLMs' performance before and after \approach transformations. All numbers are percentages (\%). B and I after model names indicate base and instruction-tuned, respectively. Top four models with the largest and smallest performance drops ($\Delta$) are highlighted in \colorbox[HTML]{F4CCCC}{red} and \colorbox[HTML]{D9EAD3}{green}. \textcolor{textgreen}{Green} and \textcolor{textred}{red} indicate whether adding a \approach ICL results in less performance drop. Same model families are grouped together.}

\label{tab:model-performance}
\vspace{-8pt}
\resizebox{\textwidth}{!}{
\begin{tabular}{|cc|cccc|cccc|cccc|cccc|cccc|c|c|c|}
\hline
\multicolumn{2}{|c|}{} & \multicolumn{4}{c|}{\textbf{Input Prediction}} & \multicolumn{4}{c|}{\textbf{Output Prediction}} & \multicolumn{4}{c|}{\textbf{Code Translation}} & \multicolumn{4}{c|}{\textbf{Program Repair}} & \multicolumn{4}{c|}{\textbf{Program Repair}} &  &  &  \\ \cline{3-22}
\multicolumn{2}{|c|}{} & \multicolumn{4}{c|}{\textbf{CRUXEval}} & \multicolumn{4}{c|}{\textbf{CRUXEval}} & \multicolumn{4}{c|}{\textbf{Avatar}} & \multicolumn{4}{c|}{\textbf{HumanEval}} & \multicolumn{4}{c|}{\textbf{ClassEval}} &  &  &  \\ \cline{3-22}
\multicolumn{2}{|c|}{\multirow{-3}{*}{\textbf{Model}}} & \multicolumn{1}{c|}{Bef.} & \multicolumn{1}{c|}{Aft.} & \multicolumn{1}{c|}{$\Delta$} & $\Delta$(IC) & \multicolumn{1}{c|}{Bef.} & \multicolumn{1}{c|}{Aft.} & \multicolumn{1}{c|}{$\Delta$} & $\Delta$(IC) & \multicolumn{1}{c|}{Bef.} & \multicolumn{1}{c|}{Aft.} & \multicolumn{1}{c|}{$\Delta$} & $\Delta$(IC) & \multicolumn{1}{c|}{Bef.} & \multicolumn{1}{c|}{Aft.} & \multicolumn{1}{c|}{$\Delta$} & $\Delta$(IC) & \multicolumn{1}{c|}{Bef.} & \multicolumn{1}{c|}{Aft.} & \multicolumn{1}{c|}{$\Delta$} & $\Delta$ (IC) & \multirow{-3}{*}{\textbf{Avg -$\Delta$}} & \multirow{-3}{*}{\textbf{Avg-$\Delta$(IC)}} & \multirow{-3}{*}{\textbf{SWE-Sub}} \\ \hline
\multicolumn{1}{|c|}{1} & CodeLlama-13-B & \multicolumn{1}{c|}{49.0} & \multicolumn{1}{c|}{42.5} & \multicolumn{1}{c|}{\cellcolor[HTML]{F4CCCC}-13.3} & {\color[HTML]{38761D} -12.2} & \multicolumn{1}{c|}{32.0} & \multicolumn{1}{c|}{20.0} & \multicolumn{1}{c|}{\cellcolor[HTML]{F4CCCC}-37.5} & {\color[HTML]{CC0000} -42.2} & \multicolumn{1}{c|}{4.8} & \multicolumn{1}{c|}{0.8} & \multicolumn{1}{c|}{\cellcolor[HTML]{F4CCCC}-83.3} & {\color[HTML]{CC0000} -100.0} & \multicolumn{1}{c|}{25.0} & \multicolumn{1}{c|}{8.5} & \multicolumn{1}{c|}{\cellcolor[HTML]{F4CCCC}-65.9} & {\color[HTML]{CC0000} -95.1} & \multicolumn{1}{c|}{9.0} & \multicolumn{1}{c|}{4.0} & \multicolumn{1}{c|}{\cellcolor[HTML]{F4CCCC}-55.6} & {\color[HTML]{CC0000} -100.0} & \cellcolor[HTML]{F4CCCC}-51.1 & \cellcolor[HTML]{F4CCCC}-69.9 & 12.8 \\ \hline
\multicolumn{1}{|c|}{2} & CodeLlama--13-I & \multicolumn{1}{c|}{47.0} & \multicolumn{1}{c|}{41.0} & \multicolumn{1}{c|}{-12.8} & {\color[HTML]{38761D} -7.4} & \multicolumn{1}{c|}{44.5} & \multicolumn{1}{c|}{31.5} & \multicolumn{1}{c|}{-29.2} & {\color[HTML]{38761D} -20.2} & \multicolumn{1}{c|}{24.4} & \multicolumn{1}{c|}{2.8} & \multicolumn{1}{c|}{\cellcolor[HTML]{F4CCCC}-88.5} & {\color[HTML]{38761D} -86.9} & \multicolumn{1}{c|}{51.2} & \multicolumn{1}{c|}{35.4} & \multicolumn{1}{c|}{-31.0} & {\color[HTML]{CC0000} -61.9} & \multicolumn{1}{c|}{23.0} & \multicolumn{1}{c|}{5.0} & \multicolumn{1}{c|}{\cellcolor[HTML]{F4CCCC}-78.3} & {\color[HTML]{38761D} -65.2} & \cellcolor[HTML]{F4CCCC}-47.9 & \cellcolor[HTML]{F4CCCC}-48.3 & 13.8 \\ \hline
\multicolumn{1}{|c|}{3} & CodeLlama--34-I & \multicolumn{1}{c|}{49.5} & \multicolumn{1}{c|}{46.5} & \multicolumn{1}{c|}{-6.1} & {\color[HTML]{CC0000} -19.2} & \multicolumn{1}{c|}{49.5} & \multicolumn{1}{c|}{34.0} & \multicolumn{1}{c|}{\cellcolor[HTML]{F4CCCC}-31.3} & {\color[HTML]{38761D} -26.3} & \multicolumn{1}{c|}{28.0} & \multicolumn{1}{c|}{5.2} & \multicolumn{1}{c|}{\cellcolor[HTML]{F4CCCC}-81.4} & {\color[HTML]{38761D} -72.9} & \multicolumn{1}{c|}{58.5} & \multicolumn{1}{c|}{50.0} & \multicolumn{1}{c|}{\cellcolor[HTML]{D9EAD3}-14.6} & {\color[HTML]{CC0000} -44.8} & \multicolumn{1}{c|}{22.0} & \multicolumn{1}{c|}{10.0} & \multicolumn{1}{c|}{\cellcolor[HTML]{F4CCCC}-54.5} & {\color[HTML]{CC0000} -54.5} & -37.6 & -43.5 & 17.0 \\ \hline
\multicolumn{1}{|c|}{4} & DeepSeekCoder-6.7-B & \multicolumn{1}{c|}{43.0} & \multicolumn{1}{c|}{39.5} & \multicolumn{1}{c|}{-8.1} & {\color[HTML]{38761D} -7.0} & \multicolumn{1}{c|}{43.0} & \multicolumn{1}{c|}{28.0} & \multicolumn{1}{c|}{\cellcolor[HTML]{F4CCCC}-34.9} & {\color[HTML]{38761D} -26.7} & \multicolumn{1}{c|}{24.8} & \multicolumn{1}{c|}{9.6} & \multicolumn{1}{c|}{-61.3} & {\color[HTML]{CC0000} -69.4} & \multicolumn{1}{c|}{59.1} & \multicolumn{1}{c|}{36.0} & \multicolumn{1}{c|}{-39.2} & {\color[HTML]{CC0000} -59.8} & \multicolumn{1}{c|}{26.0} & \multicolumn{1}{c|}{16.0} & \multicolumn{1}{c|}{-38.5} & {\color[HTML]{CC0000} -50.0} & -36.4 & -42.6 & 4.3 \\ \hline
\multicolumn{1}{|c|}{5} & DeepSeekCoder-6.7-I & \multicolumn{1}{c|}{47.0} & \multicolumn{1}{c|}{23.0} & \multicolumn{1}{c|}{\cellcolor[HTML]{F4CCCC}-51.1} & {\color[HTML]{38761D} -43.6} & \multicolumn{1}{c|}{45.0} & \multicolumn{1}{c|}{33.5} & \multicolumn{1}{c|}{-25.6} & {\color[HTML]{38761D} -22.2} & \multicolumn{1}{c|}{51.6} & \multicolumn{1}{c|}{14.8} & \multicolumn{1}{c|}{-71.3} & {\color[HTML]{38761D} -65.9} & \multicolumn{1}{c|}{70.1} & \multicolumn{1}{c|}{48.8} & \multicolumn{1}{c|}{-30.4} & {\color[HTML]{CC0000} -42.6} & \multicolumn{1}{c|}{42.0} & \multicolumn{1}{c|}{29.0} & \multicolumn{1}{c|}{-31.0} & {\color[HTML]{CC0000} -47.6} & -41.9 & -44.4 & 6.4 \\ \hline
\multicolumn{1}{|c|}{6} & DeepSeekCoder-33-I & \multicolumn{1}{c|}{54.0} & \multicolumn{1}{c|}{48.0} & \multicolumn{1}{c|}{-11.1} & {\color[HTML]{CC0000} -12.0} & \multicolumn{1}{c|}{58.5} & \multicolumn{1}{c|}{45.0} & \multicolumn{1}{c|}{\cellcolor[HTML]{D9EAD3}-23.1} & {\color[HTML]{CC0000} -23.1} & \multicolumn{1}{c|}{61.6} & \multicolumn{1}{c|}{30.0} & \multicolumn{1}{c|}{-51.3} & {\color[HTML]{CC0000} -64.3} & \multicolumn{1}{c|}{79.9} & \multicolumn{1}{c|}{50.6} & \multicolumn{1}{c|}{-36.6} & {\color[HTML]{CC0000} -45.0} & \multicolumn{1}{c|}{51.0} & \multicolumn{1}{c|}{38.0} & \multicolumn{1}{c|}{-25.5} & {\color[HTML]{38761D} -23.5} & \cellcolor[HTML]{D9EAD3}-29.5 & \cellcolor[HTML]{D9EAD3}-33.6 & 10.6 \\ \hline
\multicolumn{1}{|c|}{7} & DeepSeek-R1 & \multicolumn{1}{c|}{89.5} & \multicolumn{1}{c|}{84.0} & \multicolumn{1}{c|}{-6.1} & {\color[HTML]{38761D} -2.2} & \multicolumn{1}{c|}{98.5} & \multicolumn{1}{c|}{97.0} & \multicolumn{1}{c|}{\cellcolor[HTML]{D9EAD3}-1.5} & {\color[HTML]{38761D} -0.5} & \multicolumn{1}{c|}{89.6} & \multicolumn{1}{c|}{72.8} & \multicolumn{1}{c|}{\cellcolor[HTML]{D9EAD3}-18.8} & {\color[HTML]{38761D} -13.4} & \multicolumn{1}{c|}{92.7} & \multicolumn{1}{c|}{89.0} & \multicolumn{1}{c|}{\cellcolor[HTML]{D9EAD3}-3.9} & {\color[HTML]{CC0000} -7.2} & \multicolumn{1}{c|}{48.0} & \multicolumn{1}{c|}{45.0} & \multicolumn{1}{c|}{\cellcolor[HTML]{D9EAD3}-6.3} & {\color[HTML]{CC0000} -18.8} & \cellcolor[HTML]{D9EAD3}-7.3 & \cellcolor[HTML]{D9EAD3}-8.4 & 48.4 \\ \hline
\multicolumn{1}{|c|}{8} & Semcoder & \multicolumn{1}{c|}{53.5} & \multicolumn{1}{c|}{50.5} & \multicolumn{1}{c|}{\cellcolor[HTML]{D9EAD3}-5.6} & {\color[HTML]{CC0000} -8.4} & \multicolumn{1}{c|}{55.0} & \multicolumn{1}{c|}{38.5} & \multicolumn{1}{c|}{-30.0} & {\color[HTML]{38761D} -19.1} & \multicolumn{1}{c|}{44.8} & \multicolumn{1}{c|}{17.6} & \multicolumn{1}{c|}{-60.7} & {\color[HTML]{38761D} -55.4} & \multicolumn{1}{c|}{78.7} & \multicolumn{1}{c|}{50.0} & \multicolumn{1}{c|}{-36.4} & {\color[HTML]{CC0000} -62.8} & \multicolumn{1}{c|}{32.0} & \multicolumn{1}{c|}{24.0} & \multicolumn{1}{c|}{\cellcolor[HTML]{D9EAD3}-25.0} & {\color[HTML]{CC0000} -50.0} & -31.6 & -39.1 & 0.0 \\ \hline
\multicolumn{1}{|c|}{9} & StarCoder2-15 & \multicolumn{1}{c|}{51.5} & \multicolumn{1}{c|}{50.0} & \multicolumn{1}{c|}{\cellcolor[HTML]{D9EAD3}-2.9} & {\color[HTML]{CC0000} -5.8} & \multicolumn{1}{c|}{50.5} & \multicolumn{1}{c|}{33.0} & \multicolumn{1}{c|}{\cellcolor[HTML]{F4CCCC}-34.7} & {\color[HTML]{38761D} -23.8} & \multicolumn{1}{c|}{10.0} & \multicolumn{1}{c|}{5.2} & \multicolumn{1}{c|}{\cellcolor[HTML]{D9EAD3}-48.0} & {\color[HTML]{CC0000} -100.0} & \multicolumn{1}{c|}{46.3} & \multicolumn{1}{c|}{16.5} & \multicolumn{1}{c|}{\cellcolor[HTML]{F4CCCC}-64.5} & {\color[HTML]{CC0000} -93.4} & \multicolumn{1}{c|}{19.0} & \multicolumn{1}{c|}{5.0} & \multicolumn{1}{c|}{\cellcolor[HTML]{F4CCCC}-73.7} & {\color[HTML]{CC0000} -78.9} & -44.7 & \cellcolor[HTML]{F4CCCC}-60.4 & 1.1 \\ \hline
\multicolumn{1}{|c|}{10} & WizardCoder-15 & \multicolumn{1}{c|}{34.0} & \multicolumn{1}{c|}{20.0} & \multicolumn{1}{c|}{\cellcolor[HTML]{F4CCCC}-41.2} & {\color[HTML]{CC0000} -66.2} & \multicolumn{1}{c|}{37.0} & \multicolumn{1}{c|}{26.5} & \multicolumn{1}{c|}{-28.4} & {\color[HTML]{38761D} -20.3} & \multicolumn{1}{c|}{33.6} & \multicolumn{1}{c|}{7.2} & \multicolumn{1}{c|}{\cellcolor[HTML]{F4CCCC}-78.6} & {\color[HTML]{38761D} -77.4} & \multicolumn{1}{c|}{59.1} & \multicolumn{1}{c|}{32.3} & \multicolumn{1}{c|}{\cellcolor[HTML]{F4CCCC}-45.4} & {\color[HTML]{CC0000} -75.3} & \multicolumn{1}{c|}{31.0} & \multicolumn{1}{c|}{15.0} & \multicolumn{1}{c|}{-51.6} & {\color[HTML]{CC0000} -87.1} & \cellcolor[HTML]{F4CCCC}-49.0 & \cellcolor[HTML]{F4CCCC}-65.2 & 0.0 \\ \hline
\multicolumn{1}{|c|}{11} & WizardCoder-33 & \multicolumn{1}{c|}{40.5} & \multicolumn{1}{c|}{30.0} & \multicolumn{1}{c|}{\cellcolor[HTML]{F4CCCC}-25.9} & {\color[HTML]{38761D} -18.5} & \multicolumn{1}{c|}{42.5} & \multicolumn{1}{c|}{32.5} & \multicolumn{1}{c|}{-23.5} & {\color[HTML]{38761D} -22.4} & \multicolumn{1}{c|}{32.8} & \multicolumn{1}{c|}{7.2} & \multicolumn{1}{c|}{-78.0} & {\color[HTML]{CC0000} -85.4} & \multicolumn{1}{c|}{63.4} & \multicolumn{1}{c|}{21.3} & \multicolumn{1}{c|}{\cellcolor[HTML]{F4CCCC}-66.3} & {\color[HTML]{CC0000} -63.5} & \multicolumn{1}{c|}{31.0} & \multicolumn{1}{c|}{21.0} & \multicolumn{1}{c|}{\cellcolor[HTML]{D9EAD3}-32.3} & {\color[HTML]{CC0000} -51.6} & \cellcolor[HTML]{F4CCCC}-45.2 & \cellcolor[HTML]{F4CCCC}-48.3 & 7.4 \\ \hline
\multicolumn{1}{|c|}{12} & GPT-4o & \multicolumn{1}{c|}{79.0} & \multicolumn{1}{c|}{75.5} & \multicolumn{1}{c|}{\cellcolor[HTML]{D9EAD3}-4.4} & {\color[HTML]{CC0000} -20.9} & \multicolumn{1}{c|}{88.5} & \multicolumn{1}{c|}{84.5} & \multicolumn{1}{c|}{\cellcolor[HTML]{D9EAD3}-4.5} & {\color[HTML]{CC0000} -21.5} & \multicolumn{1}{c|}{73.2} & \multicolumn{1}{c|}{41.2} & \multicolumn{1}{c|}{\cellcolor[HTML]{D9EAD3}-43.7} & {\color[HTML]{38761D} -38.3} & \multicolumn{1}{c|}{83.5} & \multicolumn{1}{c|}{76.8} & \multicolumn{1}{c|}{\cellcolor[HTML]{D9EAD3}-8.0} & {\color[HTML]{CC0000} -10.2} & \multicolumn{1}{c|}{38.0} & \multicolumn{1}{c|}{21.0} & \multicolumn{1}{c|}{-44.7} & {\color[HTML]{38761D} 15.8} & \cellcolor[HTML]{D9EAD3}-21.1 & \cellcolor[HTML]{D9EAD3}-15.0 & 47.9 \\ \hline
\multicolumn{1}{|c|}{13} & o4-mini & \multicolumn{1}{c|}{99.0} & \multicolumn{1}{c|}{94.5} & \multicolumn{1}{c|}{\cellcolor[HTML]{D9EAD3}-4.5} & {\color[HTML]{38761D} -2.5} & \multicolumn{1}{c|}{99.0} & \multicolumn{1}{c|}{96.0} & \multicolumn{1}{c|}{\cellcolor[HTML]{D9EAD3}-3.0} & {\color[HTML]{CC0000} -4.0} & \multicolumn{1}{c|}{86.0} & \multicolumn{1}{c|}{68.0} & \multicolumn{1}{c|}{\cellcolor[HTML]{D9EAD3}-20.9} & {\color[HTML]{38761D} -18.1} & \multicolumn{1}{c|}{95.7} & \multicolumn{1}{c|}{73.2} & \multicolumn{1}{c|}{\cellcolor[HTML]{D9EAD3}-23.6} & {\color[HTML]{CC0000} -15.3} & \multicolumn{1}{c|}{45.0} & \multicolumn{1}{c|}{35.0} & \multicolumn{1}{c|}{\cellcolor[HTML]{D9EAD3}-22.2} & {\color[HTML]{CC0000} -28.9} & \cellcolor[HTML]{D9EAD3}-14.9 & \cellcolor[HTML]{D9EAD3}-13.8 & 58.7 \\ \hline
\multicolumn{2}{|c|}{\textbf{Avg}} & \multicolumn{1}{c|}{56.7} & \multicolumn{1}{c|}{49.6} & \multicolumn{1}{c|}{\textbf{-14.9}} & \textbf{-17.4} & \multicolumn{1}{c|}{57.2} & \multicolumn{1}{c|}{46.2} & \multicolumn{1}{c|}{\textbf{-23.6}} & \textbf{-20.9} & \multicolumn{1}{c|}{43.5} & \multicolumn{1}{c|}{21.7} & \multicolumn{1}{c|}{\textbf{-60.5}} & \textbf{-65.2} & \multicolumn{1}{c|}{66.4} & \multicolumn{1}{c|}{45.3} & \multicolumn{1}{c|}{\textbf{-35.8}} & \textbf{-52.1} & \multicolumn{1}{c|}{32.1} & \multicolumn{1}{c|}{20.6} & \multicolumn{1}{c|}{\textbf{-41.5}} & \textbf{-49.3} & \textbf{-35.2} & \textbf{-41.0} & 17.6 \\ \hline
\end{tabular}
}
\end{table}

\begin{figure}[H]
    \centering
    \vspace{-8pt}
    \includegraphics[width=\textwidth]{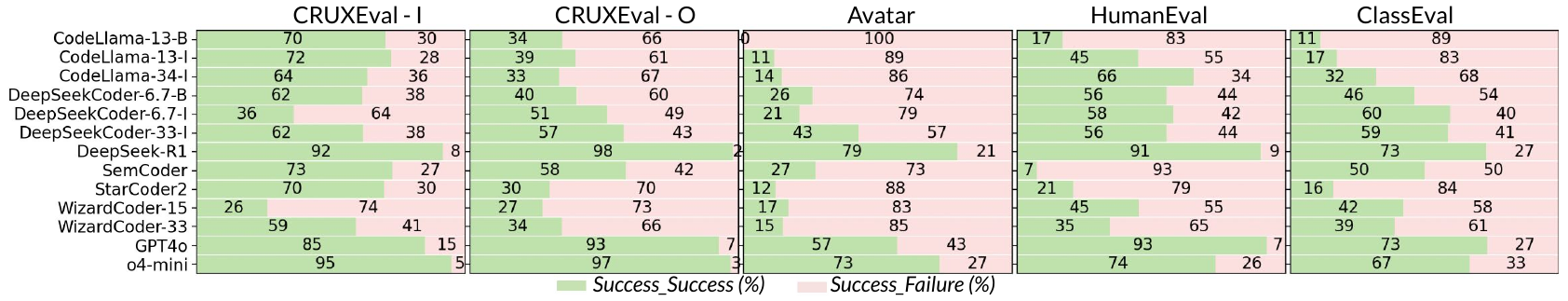}
    \vspace{-15pt}
    \caption{Performance change of LLMs on previous success after transformations}
    \label{fig:tracking}
\end{figure}

To show the effectiveness of \approach in challenging the generalizability of LLMs to more complex problems, we measured the performance of subject LLMs on selected tasks \emph{before} and \emph{after} transformations, i.e., \emph{Bef.} and \emph{Aft.} columns in Table~\ref{tab:model-performance}. Columns \textit{$\Delta$} measure percentage drop as $\Delta=(Aft-Bef)/Bef$. Figure~\ref{fig:tracking} focuses on cases across tasks that LLMs successfully solved before transformation and tracks LLMs' performance after transformation. \textcolor[HTML]{3F6A3F}{Green} and \textcolor[HTML]{B24D44}{red} bars indicate the performance of cases where LLM succeeded ($Success\_Success$) and failed ($Success\_Failure$) after transformation. These results, with a high statistical significance (p-value$=6e-18 \leq 0.05$), show that all models encounter a significant performance drop (negative $\Delta$ values) across all tasks.

Across all datasets, each original program is evolved with $10.7$ operators, on average. For those that challenged LLMs from success to failure (Figure~\ref{fig:tracking}), the mean number of transformations is $12.8$, higher than average, indicating that more transformations result in more challenging problems. To confirm this, we applied only a single operator to each problem, generating multiple applicable transformations for each, and repeated the experiment three times to account for non-determinism. Across all runs, LLMs produced almost the same outcome, confirming the impact of the combination of transformations in producing challenging problems.  

\subsubsection{Analysis Across Tasks} Concerning \textbf{code reasoning}, the average performance drop is lower in the input prediction ($14.9\%$) compared to the output prediction ($23.6\%$). We speculate this happens since models typically use one-directional forward reasoning for output predication. In contrast, input prediction requires \emph{an additional backward reasoning} step, helping LLMs understand context and deal with complexities better. 

The highest decline in average performance ($60.5\%$) among all tasks in \textbf{code translation}. We believe this is because code translation is a more complex task than others, requiring an understanding of the semantics and complexities of two programming languages~\cite{ibrahimzada2024repository}: The models should not only understand complex added features such as decorators in Python, but also find equivalence to them in the Java programming. For \textbf{program repair}, models exhibit slightly higher performance drop on \classeval ($41.5\%$) compared to \humaneval ($35.8\%$). We speculate this is because original \classeval programs are more complex than \humaneval (Figure~\ref{fig:benchmark-overview}), while transformations introduced more challenges to the code, resulting in lower average \emph{Bef.} and hence, higher $\Delta$.


\subsubsection{Analysis Across Models} \textbf{GPT-4o, o4-mini and DeepSeek-R1 have the highest number of \colorbox[HTML]{D9EAD3}{green} cells, demonstrating their leading performance.} For GPT-4o, this is likely due to the larger size and higher quality of instruction-tuning. For o4-mini and DeepSeek-R1, the strength is likely coming from the reasoning abilities of the models. \deepseek-33-I has the lowest performance drop across all open source models, on average. We speculate this is due to a large portion of code in its training data set ($87\%$) and human instruction tuning~\cite{guo2024deepseek}. \wizard-15B and \codellama-13B-Base  exhibit the most significant performance drops, followed by  \codellama-13B-Instruct, \wizard-33B, and Starcoder-2, indicating their weakness in domain transfer, i.e., their failure to capture real-world complexities from open-source programs they have trained on.

\vspace{-5pt}
\subsubsection{Analysis of Size and Training Strategies} On average, \textbf{reasoning models, instruction-tuned models, and larger models achieve higher success rates and experience smaller performance drops than base models within families}. One exception is \sem, which is instruction-tuned on \deepseek-6.7B-Base with execution data. This model exhibits less performance drop in some tasks and is even better than \deepseek-33B-Inst in input prediction, which is likely because it is instruction-tuned with dynamic execution data. 

\subsubsection{Analysis of Performance Drop.}
\label{subsubsec:embedding}

We wanted to see what factors impact the performance drop. The first speculation is a data representation shift~\cite{vaswani2017attention}. Hence, we extracted representations through mean pooling~\cite{chen2018enhancing} over the last hidden states of each token and used them to compute the \emph{Nearest Neighbor Mismatch Rate} ($NNMR$)~\cite{nnsearch}, a metric to quantify the percentage of programs whose nearest neighbor in the representation space is not their transformed counterparts.

\begin{table}[t]
\centering
\setlength{\tabcolsep}{3pt}
\vspace{10pt}
\caption{Nearest Neighbor Mismatch Rate (\textit{NNMR}) of representations. Numbers are percentages (\%).}
\vspace{-8pt}
\label{tab:nnmr}
\footnotesize
\resizebox{0.95\textwidth}{!}{
\begin{tabular}{|l|c|c|c|c|c|c|c|c|c|c|c|c|c|c|}
\hline
\textbf{Model} & D6.7-B & D6.7-I & Sem & D33-I & C13-B & C13-I & C34-I & Star & W15 & W33 & Avg.  NNMR & Avg. Token (Bef.) & Avg. Token (Aft.) \\
\hline
\textbf{CRUXEval} & 96.5 & 96.0 & 97.5 & 98.0 & 98.5 & 99.5 & 98.0 & 98.0 & 97.5 & 97.0 & 97.65 & 48.76 & 303.60 \\ \hline
\textbf{Avatar} & 77.6 & 76.8 & 77.6 & 78.8 & 80.8 & 82.4 & 79.2 & 82.0 & 74.4 & 70.8 & 78.04 & 180.66 & 467.86 \\ \hline
\textbf{HumanEval} & 37.8 & 38.4 & 35.4 & 28.1 & 43.9 & 49.4 & 44.5 & 45.7 & 32.9 & 24.4 & 38.05 & 217.11 & 458.28 \\ \hline
\textbf{ClassEval} & 18.0 & 18.0 & 17.0 & 19.0 & 25.0 & 20.0 & 22.0 & 27.0 & 13.0 & 15.0 & 19.40 & 400.48 & 688.37 \\ \hline
\end{tabular}%
}
\vspace{-10pt}
\end{table}

The results in Table~\ref{tab:nnmr} demonstrate a distribution shift, and the higher the percentage of newly added tokens, the bigger the $NNMR$ rate. For example, tokens in \cruxeval transformations are $6.2$ times that of the original programs, resulting in $97.65\%$ $NNMR$. The token growth for \classeval transformations are less ($1.7$ times), and hence, representation shift is moderate ($19.40\%$). At the same time, there is no correlation between average performance drop values (last row of Table~\ref{tab:model-performance} under $\Delta$ columns) and severity of representation shift. For example, despite the highest data representation shift, \cruxeval-I has the lowest $\Delta$. Similarly, $\Delta$ for \classeval is higher than \cruxeval and close to \humaneval, despite the lowest $NNMR$. These results suggest that \textbf{although transformations show inevitable representation shift, there might be another reason for the performance drop.} 

\begin{wrapfigure}{r}{0.5\columnwidth}
\vspace{-10pt}
\includegraphics[width=0.5\columnwidth]{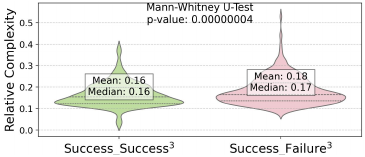}
    \vspace{-18pt}
    \caption{Comparison between relative complexity of program in $Success\_Success^3$ and $Succes\_Failure^3$}
    \vspace{-4pt}
    \label{fig:significance-test}
\end{wrapfigure}

To investigate further, we focused on $Success\_Success$ and $Success\_Failure$ programs (Figure~\ref{fig:tracking}). To minimize the impact of randomness, we constructed $Success\_Success^3$ and $Success\_Failure^3$ by focusing on <program, transformation> pairs that appear in at least \textit{three} models' $Success\_Success$ or $Success\_Failure$. We also removed the overlap between the two groups. Figure~\ref{fig:significance-test} shows the distribution of the relative complexity (Equation~\ref{eq:relative-comp}) for the transformations in the selected groups. The p-value of $4e-8 \leq 0.05$ indicates a statistically significant difference between the relative complexity distribution of transformations in two groups, \textbf{confirming the success of \approach in challenging LLMs concerning the complexity of programming problems.}

The results in Table~\ref{tab:nnmr} also rule out the potential impact of increased input length on LLMs' performance, and hence, exhausting the context window: the average token length of \approach-\humaneval (column Avg. Token (Aft.)) is close to original \classeval (column Avg. Token (Bef.)), and both are negligible compared to context window sizes of studied models (at least $8k$). 
\subsection{RQ3: Efficacy}
\label{subsec:rq3}

We further investigate whether \approach can serve as a long-time programming benchmark without risk of benchmark saturation~\cite{ribeiro2020beyond}, overfitting, or data contamination. 
\approach programs are immune to data contamination due to scalability and evolution: \approach applies a nondeterministic combination of operators to evolve programs to their more complex version (Algorithm~\ref{alg:mutation}). As we will show (\S \ref{subsub:repeat}), users can repeat evolution for a new complex version of programs than what was used in our experiments, observing a similar performance drop. Even within one evolution attempt, users can choose any of the Pareto-optimal solutions (green instances in Figure~\ref{fig:front}) rather than the most complex ones. However, \textbf{LLMs are perfect in pattern matching, raising the question of whether fine-tuning or few-shot learning based on existing \approach programs helps them bypass the introduced complexity with inflated performance on newly generated or other instances of \approach programs}. 

\subsubsection{Few-shot Learning.} For this experiment, we updated the prompt templates in Figure~\ref{fig-templates} with in-context examples from \approach: for code translation and code reasoning, we selected a Pareto-optimal solution other than the query problem used in the previous experiment to instruct the model. We could not include transformations of the same program for bug repair due to solution leakage. Instead, we selected a program in the original benchmark with the highest embedding similarity to the original version of the query program. We then used one of its Pareto-optimal solutions as an in-context example.
To further \emph{help the model} and demonstrate the diversity and quality of \approach transformations, the instruction for code reasoning and translation \emph{explicitly} mentions that the in-context example is \emph{semantically equivalent} to the query problem. This was not possible to bug repair, as we chose different problems for few-shot learning as explained before. 


Columns $\Delta(IC)$ in Table~\ref{tab:model-performance} present the results of this experiment. \textbf{Even when instructed with in-context examples generated by \approach, models struggle with these tasks and experience a performance drop (negative $\Delta(IC)$ values). Except for a few cases (highlighted in green), the performance drop is even higher than in the original experiment. We believe this is another indicator that \approach generates programs that are harder to analyze and understand by LLMs concerning different programming tasks.} 

\begin{wrapfigure}{r}{0.5\columnwidth}
\vspace{-15pt}
\includegraphics[width=0.5\columnwidth]{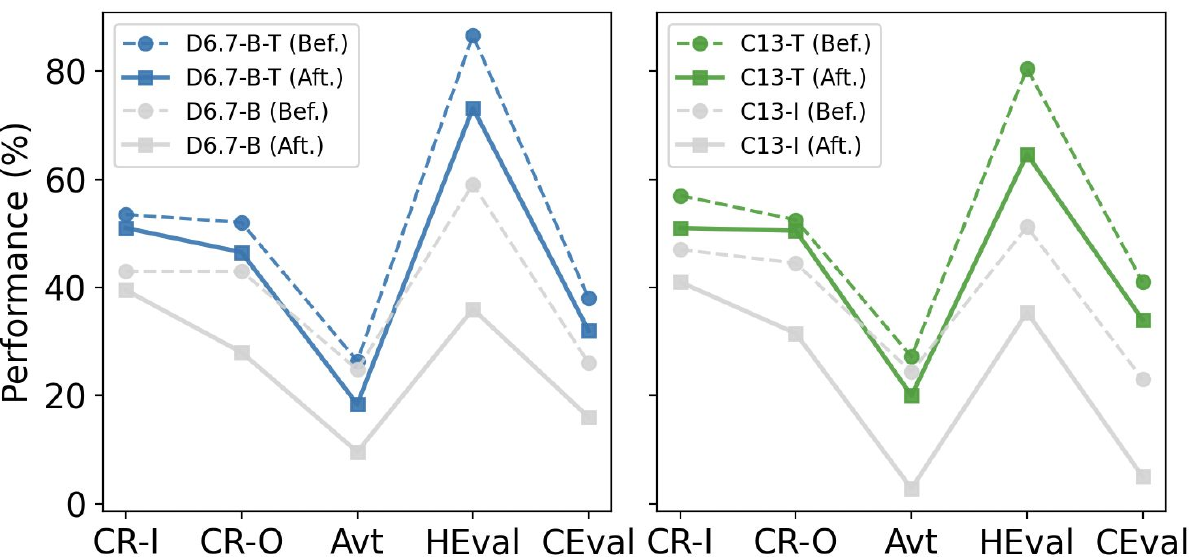}
    \vspace{-15pt}
    \caption{Model performance after fine-tuning}
    \label{fig:finetune}
\end{wrapfigure}
\subsubsection{Fine-tuning.} For this experiment, we fine-tuned CodeLlama-13-I and DeepSeekCoder-6.7-B\footnote{These models are the smallest in their respective families that still achieve competitive performance among models of similar size, while fitting within our available GPU resources for fine-tuning.}. 
To construct a fine-tuning dataset, we started with all the generated \approach transformations from all four datasets, except those that were selected to appear in the final benchmark used in previous experiments. The rationale behind selecting all transformations for fine-tuning, and not only those on the Pareto fronts, is two-fold: first of all, fine-tuning requires a substantial amount of data to work properly. Furthermore, including samples from different complexity levels enables curriculum learning, which is shown to be an effective training approach. Since the genetic algorithm can produce offspring with highly similar transformations (e.g., through variable renaming), we applied deduplication to mitigate memorization of near-duplicate programs, meanwhile reducing training costs and improving efficiency. Following best practices in LLM training~\cite{zhao20247b,lee2021deduplicating,li2023starcoder}, we filtered out samples with Jaccard similarity~\cite{broder1998min} greater than $0.8$ for each instance. This process resulted in a more diverse and balanced dataset of $106k$ fine-tuning samples.


We fine-tuned the models using QLoRA~\cite{dettmers2023qlora} on \emph{eight} $A100$-40GB GPUs for at most two epochs, applying loss-based early stopping once training plateaued. 
Figure~\ref{fig:finetune} reports the evaluation results of the fine-tuned models across all tasks under \approach benchmarks. We observe that while models show improved performance after fine-tuning, a substantial performance gap remains, ranging from {$4.7$\%} to {$30.3$\%}, with an average of $17\%$. \textbf{These findings further confirm the effectiveness of \approach in challenging the generalizability of LLMs concerning real-world programming complexities, even when fine-tuned with similar transformations.} 
\begin{wrapfigure}{r}{0.5\columnwidth}
    \vspace{3pt}
\includegraphics[width=0.45\columnwidth]{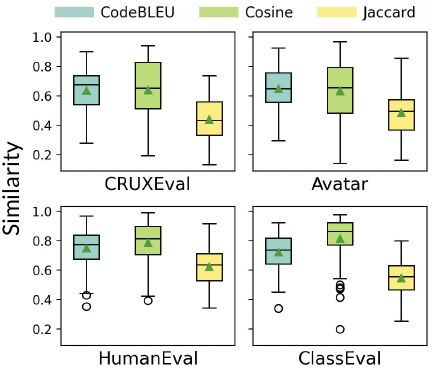}
    \vspace{-10pt}
    \caption{Similarity of three versions \approach}
    \label{fig:similarity}
\end{wrapfigure}

\vspace{-10pt}
\subsubsection{Multiple Versions.}
\label{subsub:repeat}

To assess the reproducibility of prior observations for new versions of \approach, and more specifically, to evaluate whether a similar performance drop trend is observed under new versions, we repeated the genetic algorithm process two additional times, using a unique but different seed for each generation. The average relative complexity and readability across all three versions of \approach are $0.21$ and $0.66$, respectively, close to the numbers reported in \S \ref{subsec:rq1}. 

\begin{figure*}[t]
    \centering
    \includegraphics[width=0.95\textwidth]{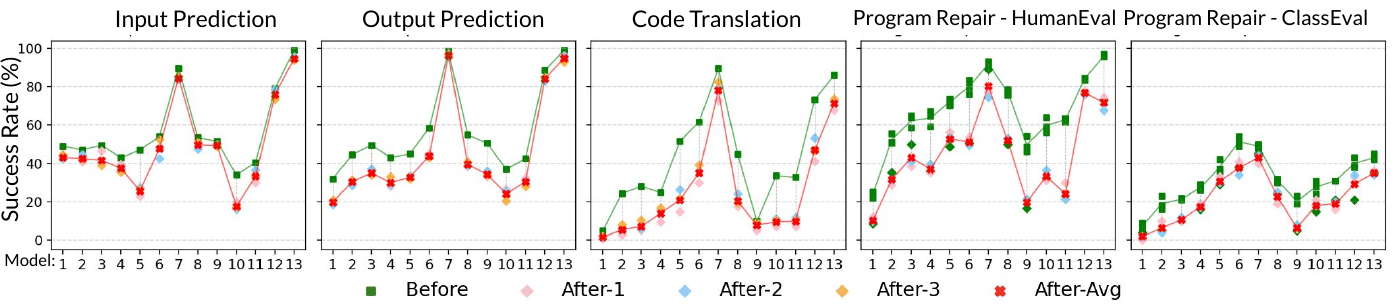}
    \vspace{-8pt}
    \caption{Performance of LLMs under three different versions of \approach. After-1 represents the original version of \approach used in previous experiments}
    \label{fig:three-times-results}
    \vspace{-10pt}
\end{figure*}

We further measured the pairwise similarity between the final \approach instances of a given program from different versions, using CodeBLEU~\cite{ren2020codebleu}, Cosine similarity~\cite{chowdhury2010introduction}, and Jaccard similarity~\cite{broder1998min} following prior work on code similarity evaluation~\cite{li2023starcoder, zhou2024evolutionary}. These metrics measure complementary aspects of similarity, including code syntax, semantics, token distribution, and token overlap.
Figure~\ref{fig:similarity} shows the result of this analysis: CodeBLEU and Cosine similarity scores mostly range between $0.60$ and $0.80$, while Jaccard scores are lower, ranging from $0.40$ to $0.64$, on average. Considering a consistent trend measured by different similarity metrics, these results indicate that \textbf{the generated programs from different runs are notably different and demonstrate non-trivial variations}. More importantly, as shown in Figure~\ref{fig:three-times-results}, all versions consistently challenge the subject LLMs on the studied downstream tasks\footnote{The performance drop, although slightly, is even higher in the After-2 and After-3 versions, mitigating the threat of reporting the best results in the paper.}, i.e., exhibiting similar performance drop trends. \textbf{These results confirm that the effectiveness of \approach is not random, reflecting its robustness and general applicability.}
\vspace{10pt}



\subsection{RQ4: Real-world Representativeness}
\label{subsec:rq4}

Finally, we investigated whether LLM's performance on \approach can be used as a proxy for its performance in the presence of real-world complexities, removing the need for constructing task-specific real-world benchmarks and dealing with corresponding limitations (\S \ref{sec:introduction}). 
This experiment required real-world benchmarks for program repair, code translation, and code reasoning. To our knowledge, no existing benchmarks are available for real-world code reasoning. For code translation, we only found RepoTransBench~\cite{wang2024repotransbench} as a repository of GitHub Python projects and their confirmed Java translations. However, we faced several non-trivial issues with the artifacts and received no response from the authors. For program repair, we chose SWE-Bench-Lite~\cite{swebenchlite}, a subset of $300$ SWE-Bench instances focused on \emph{functional bug fixes}\footnote{SWE-Bench problems can be bug fixes or feature requests. Some bug-fixing problems concern non-functional bugs.}. 

We filtered SWE-Bench-Lite instances with more than three lines of additions compared to deletions\footnote{The total number of additions/deletion/changes can be more than three.}. This ensures a fair comparison between our injected bugs and theirs to perform a controlled experiment\footnote{this paper focuses on code complexity (i.e., contextual complexity), and we plan to extend \approach to generate more complex artifacts in future.}. Filtering resulted in $94$ instances, which we refer to as \textit{SWE-Sub}. We used Agentless~\cite{agentless} as the repair tool with subject LLMs and measured its performance in fixing SWE-Sub problems. SWE-Sub requires Agentless to analyze the issue report to localize the bug, which could be at the file, class, function, or statement level. Given that LLMs already know the file-level location of the changes in \S \ref{subsec:rq1}, we skipped the Agentless file localization step. For selected buggy files, we keep the implementation of buggy classes with all methods (not just the buggy method) and pass it to Agentless for repair. The problems are still hard, as LLMs need to localize the bug to 
functions and statements.  Figure~\ref{fig:after-transformation}e shows the complexity of SWE-Sub problems. 

Comparison of the LLMs' performance in program repair (Table~\ref{tab:model-performance}) using \approach (\emph{Aft.} columns under \humaneval and \classeval) and SWE-Sub (last column) problems demonstrates notable similarities. \textbf{The average repair rates under \approach (\humaneval=$45.3\%$, \classeval=$20.6\%$) are much closer to SWE-Sub ($17.6\%$), compared to original \humaneval ($66.4\%$) and \classeval ($32.1\%$)}. \sem and \wizard-15B fail to generate valid patches, likely due to a shorter context window ($8,192$ tokens), i.e., they mostly generated a random, meaningless sequence of characters rather than valid code, respectively.

\section{Related Work}
\label{sec:relatedwork}

Several benchmarks have been constructed to evaluate LLMs' abilities on different programming tasks. \humaneval~\cite{chen2021evaluating} and \classeval~\cite{du2023classeval} assess the program synthesis abilities of LLMs at the method and class level. \avatar~\cite{ahmad2021avatar} is utilized for code translation. \cruxeval~\cite{gu2024cruxeval} is a syntactic dataset for code reasoning. \emph{These benchmarks do not represent real-world complexities (\S \ref{sec:motivation}).} 
Recently, there have been benchmarks from the GitHub repositories: SWE-Bench~\cite{jimenez2024swebench} is curated from GitHub issues. R2E~\cite{jain2024r2e} transforms existing GitHub repositories into testing environments for LLMs. RepoTransBench~\cite{wang2024repotransbench} is a collection of Python programs and their equivalent Java translations. \emph{Compared to \approach, these benchmarks suffer from specific-task evaluation, inefficient quality control, data contamination, and overfitting (\S \ref{sec:introduction}).}
Several related works attempted to transform existing \emph{code synthesis} benchmarks to new problems. PPM~\cite{chen2024ppm} transforms natural language statements in program synthesis problems to generate new problems. EvoEval~\cite{evoeval} prompts LLMs to generate problems of different difficulty levels by combining existing benchmarks. \emph{Different from prior techniques, \approach focuses on \textit{code-to-code/text} tasks to automatically evolve programming problems in existing benchmarks into more complex versions.}

\approach is similar to programming adversarial attacks~\cite{yefet2020adversarial,zhou2024evolutionary,bielik2020adversarial,schuster2021you,srikant2021generating,nguyen2021adversarial,yang2022natural} 
in the sense that both aim to maintain the functionality of the original code. However, \approach aims to assess the generalizability of LLMs to real-world complexity. The key difference is that the former focuses on domain transfer, i.e., how well LLMs trained on real-world programs apply learned knowledge to similar contexts, while the latter challenges models with semantic understanding. \emph{Preserving the semantics is not essential to assess generalizability}. However, we chose it as a design decision to reuse original benchmark artifacts and support multi-task evolution. \approach operators barely overlap with that of prior work on adversarial testing, or if they do, they are more complex compared to small and simple changes (\S \ref{sec:all-operators}).

\vspace{3pt}
\section{Threats to Validity}
\label{sec:threats}
\vspace{5pt}
\noindent \textbf{External Validity.} One of the external threats is the generalizability of \approach to different programming benchmarks. 
This paper shows its effectiveness in transforming $714$ Python programs from four different diverse datasets. Our current implementation targets Python, as it dominates benchmarks like HumanEval and SWE-Bench. 
Supporting more programming languages does not require changes in algorithms or analyses and is primarily an engineering effort. 

\vspace{5pt}
\noindent \textbf{Internal Validity.} We prompted LLMs once per task but set the temperatures to zero to account for the impact of non-determinism on the conclusions' generalizability. We load models in bfloat16 precision to ensure correct model performance and prevent accuracy loss from quantization. Furthermore, our automated in-depth analysis confirms similar patterns and root causes across all models and tasks, making the observations unlikely to be random. 

\approach relies on test execution to verify functional equivalence. The original datasets already have very high test line coverage ($90.8\%$–$99.5\%$), showing that their test suites can comprehensively exercise the programs. Importantly, \approach programs maintain comparable coverage ($93.8\%$-$99.4\%$) and pass the same tests, indicating that these tests are effective at validating functional equivalence between the original and transformed code. These statistics, along with the rigorous design and application of transformation operators using AST, increase the confidence in the semantic-perverseness of the \approach programs compared to the original ones (preserve semantics by design). 
Furthermore, \approach uses the same tests for measuring LLMs' performance on different programming tasks: Correctness of translation is only evaluated using existing tests; code reasoning is assessed for an execution path demonstrated by a given test; and program repair expects patches to pass on given tests.

\vspace{5pt}
\noindent \textbf{Construct Validity.} To mitigate construct validity, \approach is built with well-vetted tools for program analysis (Python AST library and py2cfg) and genetic algorithm design (NSGA-II). 

\section{Concluding Remarks}
\label{sec:conclusion}

\approach is an automated approach that assesses the generalizability of Code LLMs to real-world complexity. It leverages a genetic algorithm to maximize the complexities of programs to the level of real-world projects while maintaining similar readability. Evaluating $13$ LLMs across four different programming tasks on \approach transformations demonstrates its effectiveness and scalability on challenging LLMs with real-world complexity. 
\approach is just the beginning of the automated generation of realistic (concerning complexity) programming problems. We plan to augment \approach with semantic-altering transformations to explore the search space of benchmark generation, challenging the domain transfer of LLMs more. 



\balance
\bibliographystyle{ACM-Reference-Format}
\bibliography{Reference}

\end{document}